\documentclass[sigconf]{acmart}

\usepackage{colortbl}
\usepackage{geometry}
\usepackage{balance}
\usepackage{subfigure}
\usepackage{booktabs} 
\usepackage{geometry}
\usepackage{tikz}
\usepackage{multirow}
\usepackage[multiple]{footmisc}
\usetikzlibrary{arrows.meta,positioning}
\usepackage[font=rm]{caption}
\DeclareCaptionType{copyrightbox}

\usepackage[inline]{enumitem}
\newlist{inlinelist}{enumerate*}{1}
\setlist*[inlinelist,1]{%
	label=(\roman*),
}
\usepackage{comment}
\usepackage[all]{nowidow}
\usepackage{balance}
\usepackage{fancyvrb}
\usepackage{amssymb}
\usepackage{pifont}
\usepackage[toc,page]{appendix}
\newcommand{\bao}[1]{\textrm{\textcolor{blue}{Bao says: #1}}}
\newcommand{\sheng}[1]{{\color{red}{#1}}}%

\definecolor{burntorange}{rgb}{0.8, 0.33, 0.0}
\newcommand{\yuan}[1]{\textrm{\textcolor{burntorange}{Yuan says: #1}}}

\newcommand{\cmark}{\ding{51}}%
\newcommand{\xmark}{\ding{55}}%

\newcommand{\myparagraph}[1]{\vspace{0.3\baselineskip}\noindent{\textbf{#1.}}~}
\newcommand{\var}[1]{\mbox{\emph{#1}}}
\newcommand{\svar}[1]{\mbox{\scriptsize\emph{#1}}}

\newcommand{\avar}[1]{\mbox{#1}}

\DeclareMathOperator*{\argmax}{arg\,max}
\DeclareMathOperator*{\argmin}{arg\,min}
\usepackage[linesnumbered,ruled,vlined,boxed,noend]{algorithm2e}
\usepackage{url}

\newtheorem{lemma1}{Lemma}

\newtheorem{definition1}{Definition}

\SetKwComment{Comment}{$\triangleright$\ }{}

\newcommand{\ra}[1]{\renewcommand{\arraystretch}{#1}}

\usepackage{epstopdf}
\usepackage{mathtools}
\usepackage{cellspace}
\usepackage{makecell}
\usepackage{float}

\usepackage{tikz}
\usetikzlibrary{tikzmark}
\usepackage{amsmath,bm}

\DeclareMathOperator{\tr}{tr}

\newcommand{\ub}{O^{\uparrow}}%
\usepackage[group-separator={,}]{siunitx}

\newcommand{\fairbus}{{\small{\textsf{CT-Bus}}}\xspace}

\newcommand{\algrule}[1][.2pt]{\par\vskip.5\baselineskip\hrule height #1\par\vskip.5\baselineskip}
\definecolor{issuecolor}{RGB}{0,166,81}

\definecolor{light-gray}{gray}{0.95}

\newcounter{cN}
\usepackage[T1]{fontenc}

\copyrightyear{2021}
\acmYear{2021}
\setcopyright{acmlicensed}\acmConference[SIGMOD '21]{Proceedings of the 2021 International Conference on Management of Data}{June 20--25, 2021}{Virtual Event, China}
\acmBooktitle{Proceedings of the 2021 International Conference on Management of Data (SIGMOD '21), June 20--25, 2021, Virtual Event, China}
\acmPrice{15.00}
\acmDOI{10.1145/3448016.3457247}
\acmISBN{978-1-4503-8343-1/21/06}

\begin{document}
\sloppy
\fancyhead{} 
\title{Public Transport Planning: When Transit Network Connectivity Meets Commuting Demand}

\author{Sheng Wang$^1$, Yuan Sun$^2$, Christopher Musco$^1$, Zhifeng Bao$^3$}
\orcid{0000-0002-5461-4281}
\affiliation{%
	\institution{$^1$New York University, $^2$Monash University, $^3$RMIT University}
}
\email{[swang, cmusco]@nyu.edu,\space\space\space\space yuan.sun@monash.edu,\space\space\space\space zhifeng.bao@rmit.edu.au}

\renewcommand{\shortauthors}{Wang et al.}
\setcounter{page}{1}

\begin{abstract}
In this paper, we make a first attempt to incorporate  both \underline{c}ommuting demand and \underline{t}ransit network connectivity in \underline{bus} route planning (\fairbus), and formulate it as a constrained optimization problem: 
\textit{planning a new bus route with $k$ edges over an existing transit network without building new bus stops to maximize a linear aggregation of commuting demand and connectivity of the transit network.}
We prove the NP-hardness of \fairbus and propose an expansion-based greedy algorithm that iteratively scans potential candidate paths in the network.
To boost the efficiency of computing the connectivity of new networks with candidate paths, we convert it to a matrix trace estimation problem and employ a Lanczos method to estimate the natural connectivity of the transit network with a guaranteed error bound.
Furthermore, we derive upper bounds on the objective values and use them to greedily select candidates for expansion. Our experiments conducted on real-world transit networks in New York City and Chicago verify the efficiency, effectiveness, and scalability of our algorithms.




\end{abstract}

\maketitle

\section{Introduction}

With the population density increasing over time \cite{I}, the gap between the demand and the supply in public transport system is becoming larger~\cite{Jiao2013}.
Enhancing the transit network with new transit routes can reduce this gap~\cite{Schiller2017,Wu2018c}, hence having the potential to bring people from private transport to public transport \cite{Liu2016,Beirao2007,Wang2015d,Wang2021}.
%
To achieve this mission, much attention has been paid to planning new routes based on emerging demands discovered from commuting records, which is also known as \textit{demand-aware route planning} \cite{Wang2020b,Liu2016,Wang2018c}.
Unfortunately, most of these studies require constructing new bus stops.
For well-covered cities like New York City (NYC), however, 
constructing new bus stops is often unnecessary and costly.\footnote{{\emph{A NYC bus network redesign}} \cite{II} is being conducted, and will eliminate 400 stops and add new routes in the Bronx \cite{III}.}
{In contrast, thanks to the connectivity of a road network, it does not incur any extra construction cost to create new edges by linking two unconnected existing stops.}
Existing studies try to meet demands without creating new stops in various ways, such as formulating it as maximal reverse k nearest neighbor trajectory queries~\cite{Wang2018c}, or a budgeted optimization problem~\cite{Liu2016}, or optimizing existing bus routes' time schedule~\cite{Mo2021}.

{Apart from meeting passengers' commuting demands alone, we argue that an ideal transit network should also be as connected and convenient as possible for passengers to transfer, i.e., the new bus route should well connect existing routes such that more transfer options can be provided.}
As reported in \cite{Wei2014a,Abdelaty2020,Zou2013}, network connectivity is an important indicator. Our empirical study also shows that a connectivity-aware route planning can help the commuters along the new route avoid up to \textbf{4.7} transfers on average in the Bronx of NYC, while a normal demand-aware planning can only avoid around \textbf{1.6} transfers (see the bold numbers in Table~\ref{tab:num}).
Unfortunately, there has not been any transit route planning work that aims to optimize the connectivity of a transit network yet.

Motivated by the above observations, we make a first attempt to define the {objective to be optimized} as a weighted sum of \textit{transit network connectivity} and \textit{commuting demand}. {It provides a flexible way to specify configurations that can meet different planning requirements \cite{Fan2006,Geisberger2010,Batz2012}}. Consequently, our optimal bus route planning problem \fairbus can be formulated as:
{\textit{given a trajectory dataset of users' commuting records and a transit network over the road network in a city, \fairbus aims to plan a new route with at most $k$ (new and existing) edges, such that the objective value is maximized}.}
After a careful study of existing connectivity measures and an evaluation over real-life transit networks (in Section~\ref{sec:liter}), we adopt \textit{natural connectivity} \cite{Estrada2000,Wu2010a,Chan2014,Chen2018d} to measure the transit network connectivity.

Optimizing the objective of \fairbus is challenging because it is a combination of two complex constrained optimization problems over graph \cite{Wangaclustering,Chan2014}.
A straightforward solution is to generate a large number of candidate paths from the graph and choose the one with the highest objective value.
Then, for every candidate, we need to compute the connectivity of the enhanced network. 
Unfortunately, it is computationally expensive to evaluate the objective function of \fairbus since the calculation of connectivity requires the computation of eigenvalues of the adjacency matrix \cite{Estrada2000,Wu2010a}.


To overcome this challenge, we convert the connectivity computation as a \emph{matrix trace estimation problem}, and employ a Lanczos method \cite{Ubaru2017,Musco2018} to estimate the natural connectivity with a bounded error. Subsequently, we derive two upper bounds on the objective values to greedily construct candidate bus routes. Furthermore, we propose a pre-computation based method that can significantly reduce the running time and meanwhile generate bus routes with competitive objective values.
{Finally, when evaluating the proposed methods, we monitor how fast the objective values converge over real-world transit networks and propose multiple metrics to measure the transfer convenience of the new transit network.}




To summarize, this paper makes the following contributions:
\begin{itemize}[leftmargin=*]
	
	
	\item We propose and formally define a novel route planning problem-\fairbus, aiming to plan a new route without constructing new bus stops, such that commuting demands are met and meanwhile the transit connectivity is improved (Section~\ref{sec:defi}).
	
	\item We prove the NP-hardness of \fairbus and propose a general algorithm by expanding, ranking, and pruning candidate paths by traversal in the network (Section~\ref{sec:fairbus}). 
	
	\item {We identify that network connectivity computation is the major efficiency bottleneck of the above algorithm, and propose to convert it into a matrix trace estimation problem and solve it approximately with a Lanczos-based method via several iterations of simple matrix multiplications. As a result, the efficiency is boosted by up to three orders of magnitudes  (Section~\ref{sec:computeconn}).}
	
	\item {We employ pre-computations on edges' connectivity increment to devise a fast connectivity estimation method, such that the calculation of each candidate path is further accelerated} (Section~\ref{sec:expopti}). 
	
	\item We conduct experiments on real-world datasets to verify that our methods can efficiently generate a new bus route that not only brings high connectivity increment but also helps connect multiple existing routes for more convenient transfers (Section~\ref{sec:exp}). 
\end{itemize}


\section{Related Work}
\label{sec:liter}


\myparagraph{Demand-Aware Route Planing}
Compared with traditional route planning via conducting passenger surveys or estimating the demand from demographic data, demand-aware route planning aims to effectively discover timely demands from commuters,
whose methodologies can be divided into two groups: 1) network (re-)design; and 2) specific route optimization or adding a new route. 
{Specifically, \citet{Chen2014a} exploited overnight taxi trajectories to first detect the areas containing frequent pick-up/drop-offs, and then partition them into a number of clusters and identify a location in each cluster as a candidate bus stop, and finally decide popular routes from all candidate bus routes via two heuristic methods.} 
\citet{Pinelli2016} redesigned the whole transit network based on mobile phone trajectories from cell towers, 
by deriving frequent movement patterns and planning new routes in existing stops, without the consideration of transfer and connectivity.

Instead of knocking down existing transit systems, 
\citet{Liu2016} proposed to discover the routes that are not well operated, 
and optimized them based on popular origin-destination pairs extracted from taxi and bus trip records, 
by optimizing a route in the existing transit network.
Reverse k Nearest Neighbors over Trajectories (RkNNT) \cite{Wang2018c} is a tool for estimating the demand for a bus route based on trajectory data in the existing transit network, and it is used to plan a route with a maximum capacity between two given stops.
Recently, a trajectory clustering method \cite{Wangaclustering} is proposed to find $k$ representative paths (i.e., traffic trend) in the road network rather than the transit network, hence new stops need to be constructed.
To summarize, 
none of them considered whether new edges should be linked to form a route and make the network more connected like \fairbus.

\begin{figure}
	\centering
	\vspace{-1.3em}
	\includegraphics[height=2.9cm]{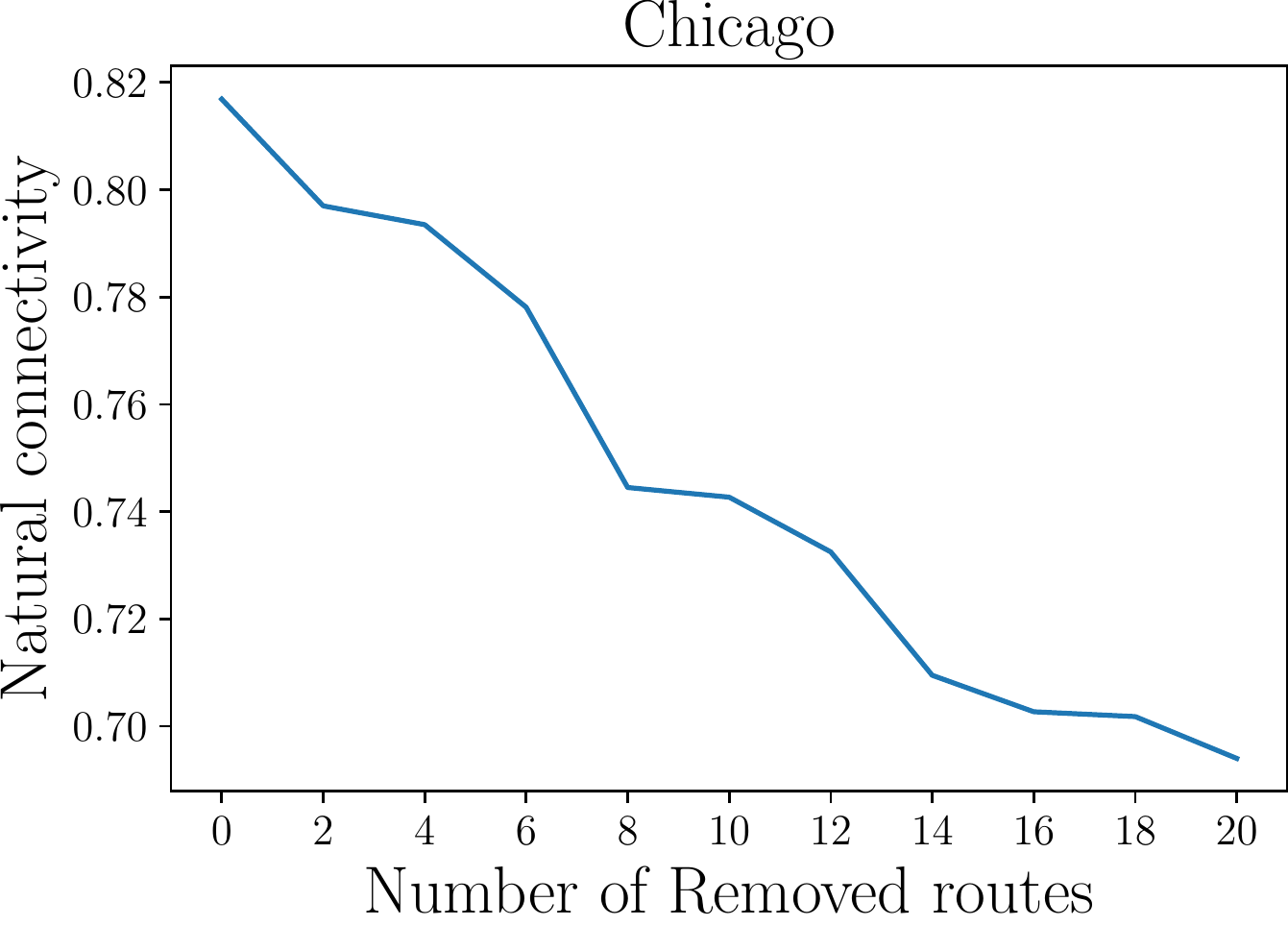}\hspace{0.5em}
	\includegraphics[height=2.9cm]{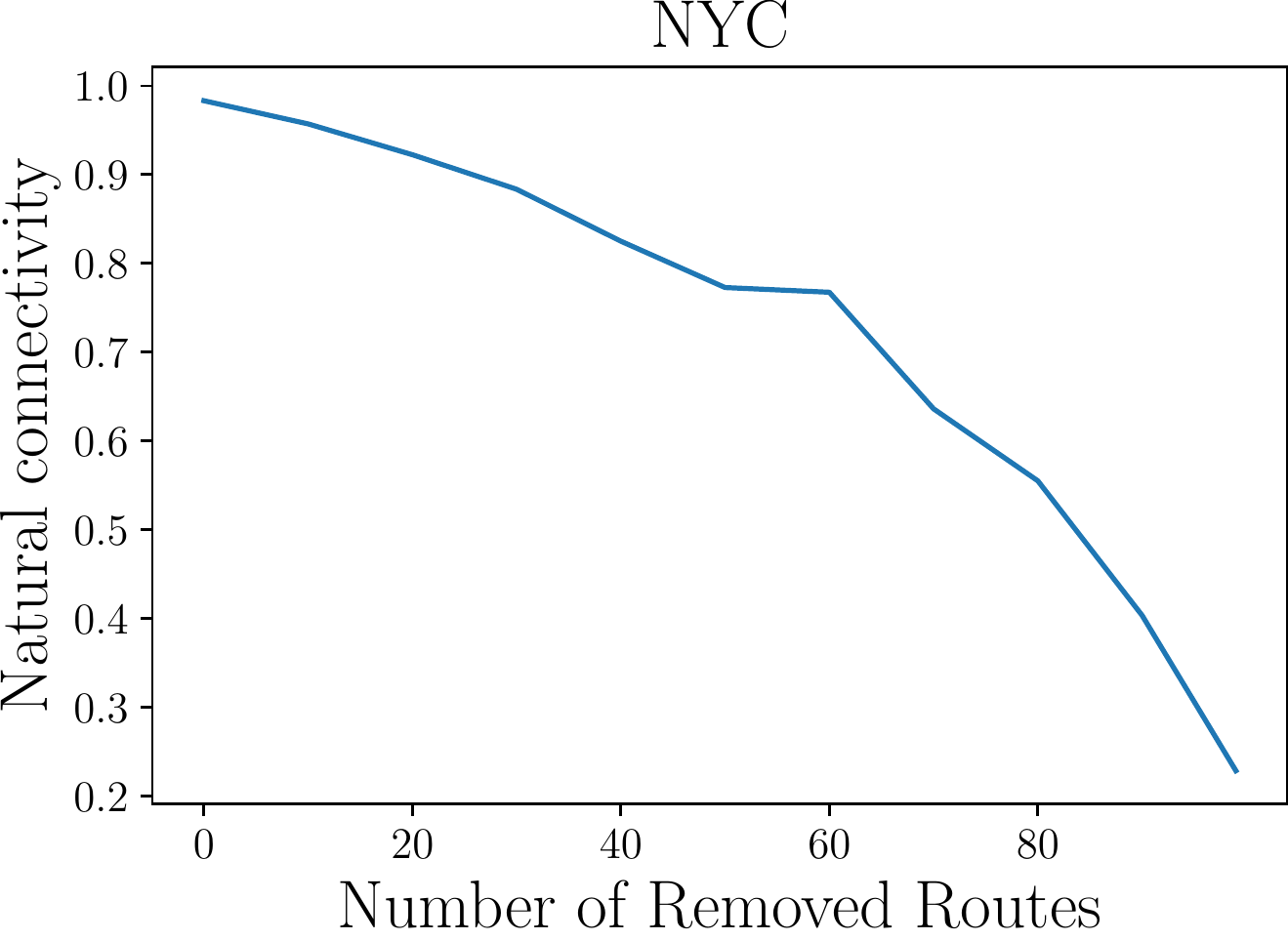}
	\vspace{-1em}
	\caption{Evaluation of natural connectivity on two real-world datasets.}
	\vspace{-2em}
	\label{fig:connectivity}
\end{figure}

\myparagraph{{Transit Network Connectivity}} 
As one of the key metrics in measuring the transfer convenience of a transit network~\cite{Zou2013,Abdelaty2020,Guihaire2008}, \emph{connectivity}~\cite{Kaplan2014} has been proposed and studied extensively in the transportation area. By modeling the transit network as an undirected graph composed of vertexes and edges, the connectivity can be measured in numerous ways, such as the vertex and edge connectivity \cite{west1996introduction},  algebraic connectivity \cite{Fiedler1972,Wei2014a},  and natural connectivity \cite{Wu2010a} (also known as an extension from \textit{Estrada index} \cite{Estrada2000}) that is more proper for complex networks. 
Among these measures, natural connectivity (defined in Equation~\ref{equ:conn}) is arguably the most proper one for transit network, since it will not show drastic changes by small graph alterations (algebraic connectivity) or no change by big graph alteration (edge connectivity); instead, it can monotonically evolve w.r.t. more modifications \cite{Chan2014}.
To verify its monotonicity in real transit networks, we randomly remove existing routes from Chicago and New York City transit networks gradually, and we observe a nearly linear decrease of natural connectivity (see Figure~\ref{fig:connectivity}). 

To our best knowledge, there has not been any study that can plan a new route to optimize the connectivity of public transport networks yet, despite the choice of connectivity adopted. {One loosely related work is \cite{Wei2014a}, aiming to optimize air transport network's connectivity by adding $k$ new discrete edges, which is also a classical graph augmentation problem \cite{Chan2014}}. However, such an edge is not a route and hence cannot be used to solve our problem; also, it does not consider the demand from commuters.
On the other side, in the graph mining field, there have been techniques aiming to improve the connectivity of a network via edge augmentation \cite{Nutov2005a,Chan2014,Chen2018d}, but the resulted edges are typically discrete and hence cannot be directly applied to plan a connected route.

\myparagraph{\textit{\underline{Comparable Approaches}}}{We call the aforementioned work \cite{Chan2014,Wei2014a} as the  \textit{connectivity-first} approach, and will use it as a baseline in our experiments. Notably, our results (see Figure~\ref{fig:baseline-topk}) show that the discrete edges are hard to be connected as a smooth bus route. Similarly, meeting the commuting demand maximally with a single bus route can be another baseline, and we call it as the  \textit{demand-first} approach; essentially, it is equivalent to the refinement step in trajectory clustering \cite{Wangaclustering} and can be implemented with proper parameter setting on our objective function formally defined in Definition~\ref{def:fairbus}, which will also be compared.} 

\section{{PROBLEM FORMULATION}}
\label{sec:defi}
\subsection{Preliminaries}

\begin{definition1}(\textbf{Road Network})
\label{def:roadnet}
A road network is an undirected graph $G=(V, E)$: $V$ is a set of
vertices representing the intersections and terminal points of
the road segments; $E$ is a set of edges representing road
segments. Vertices are indexed from $1$ to $|V|$: $\{v_1, v_2,\cdots, v_{|V|}\}$.
\end{definition1}

\begin{definition1}(\textbf{Transit Network})
	\label{def:transit}
	A transit network is an undirected graph $G_r=(V_r, E_r)$: $V_r$ is a set of vertices representing bus stops; $E_r$ is a set of edges connecting two vertices. Each vertex in $V_r$ is affiliated with an edge in $G$. Each edge $e$ corresponds to a path composed of connected edges in $G$ and $|e|$ is the travel length of edge $e$. A bus route is composed of a set of connected edges in $G_r$. 
\end{definition1}

\begin{definition1}(\textbf{Trajectory Data} \cite{Wangaclustering,Wangsheng2018})
	\label{def:maptra}
	A trajectory ${T}$ in the road network is a set of connected vertices with timestamps (the time entering each vertex) in $G$, such that
	${T}:(v_1,t_1) \rightarrow (v_2,t_2)\rightarrow \ldots\rightarrow (v_{l}, t_l)$. Each trajectory can be converted to a path in $G$ and $G_r$. 
\end{definition1}

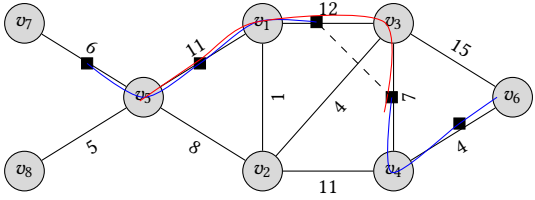
\begin{figure}
	\centering
\scalebox{1}{\begin{tikzpicture}[
      mycircle/.style={
         circle,
         draw=black,
         fill=gray,
         fill opacity = 0.3,
         text opacity=1,
         inner sep=0pt,
         minimum size=15pt,
         font=\small},
      myarrow/.style={-},
      node distance=0.6cm and 1.2cm
      ]
      
      \node[mycircle] (c1) {$v_5$};
      \node[mycircle,below right=of c1] (c2) {$v_2$};
      \node[mycircle,right=of c2] (c3) {$v_4$};
      \node[mycircle,above right=of c1] (c4) {$v_1$};
      \node[mycircle,right=of c4] (c5) {$v_3$};
      \node[mycircle,below right=of c5] (c6) {$v_6$};
      
      \node[mycircle,above left=of c1] (b1) {$v_7$};
      \node[mycircle,below left=of c1] (b2) {$v_8$};

    \foreach \i/\j/\txt/\p in {
    	 b2/c1/5/below,
	   b1/c1/6/above,
      c1/c2/8/below,
      c1/c4/11/above,
      c2/c3/11/below,
      c3/c6/4/below,
      c4/c5/12/above,
      c5/c6/15/above,
      c5/c2/4/below,
      c3/c5/7/below,
      c2/c4/1/below}
       \draw [myarrow] (\i) -- node[sloped,font=\small,\p] {\txt} (\j);
     
     \node[draw,fill=black, scale=0.7] at (-0.75,0.45) {};
     \node[draw,fill=black, scale=0.7] at (0.75,0.45) {};
     \node[draw,fill=black, scale=0.7] at (2.3,1) {};
     \node[draw,fill=black, scale=0.7] at (3.3,0) {};
     \node[draw,fill=black, scale=0.7] at (4.2,-0.35) {};
     
     \draw [blue] plot [smooth] coordinates {(3.3,0) (3.3, -1) (4.2, -0.35) (4.7, 0)};
     
     \draw [blue] plot [smooth] coordinates {(-0.75,0.45) (0,0) (0.75,0.45) (1.5,1) (2.3,1)};
     \draw [red] plot [smooth ] coordinates {(-0.05,-0.05) (0.75,0.5) (1.5,1) (3.2, 1) (3.2, -0.2)};
     \draw[dashed] [black] plot [smooth] coordinates {(3.3,0)  (2.3,1)};
\end{tikzpicture}}
	\vspace{-1em}
	\caption{An example of the road network (gray circles represent vertexes, the number denotes the edge demand), transit network (black squares indicate stops; two lines with blue color denote bus routes), and a trajectory (the red line).}
	\label{fig:example}
\end{figure}

It is worth mentioning that a raw GPS-sampled trajectory can be projected to the road network effectively via \textit{map-matching} \cite{Lou2009} with high analytic precision \cite{Wangsheng2018}.
We define both networks as undirected since bus routes are usually round trips and can be modeled into one graph. 
Figure~\ref{fig:example} is designed to illustrate all the above definitions.
The dashed line shows that a new edge is necessary for transfer between two bus routes. 
We plot the trajectory and bus routes with a little shift to roads for better visual distinction.


\subsection{Problem Definition}
\subsubsection{Transit Connectivity Measure}
Connecting nodes and generating new edges to make the network more connected can offer more transfer choices to passengers.
As discussed in Section~\ref{sec:liter}, we choose to use \textit{natural connectivity} \cite{Estrada2000,Wu2010a,Chan2014,Chen2018d} (Equation~\ref{equ:conn}). 

\begin{table}
	\centering
	\caption{Summary of major notations.}
	\vspace{-1em}
	\scalebox{0.9}{\begin{tabular}{cc}
		\toprule
		\textbf{Symbol} & \textbf{Description} \\ \midrule
		$D$ & the dataset composed of trajectories $T$\\ \midrule
		$G_r$, $G_r^{'}$&the old and new transit networks enhanced by $\mu$ \\ \midrule
		 $\bm{A}$, $V_r$, $E_r$ & the adjacency matrix, vertex, and edge set of $G_r$ \\ \midrule
		$\lambda(G_r)$ & the connectivity of transit network $G_r$ \\ \midrule
		$\mu$ & the new route composed of edges $e$ in $G_r^{'}$\\ \midrule
		$O(\mu)$, $\ub(\mu)$ & the objective value of $\mu$ and its upper bound \\ \midrule
		$O_d(\mu)$, $O_\lambda(\mu)$ & the demand and connectivity increment \\ \midrule
		$\tau$, $tn(\mu)$ & $\mu$'s edge length threshold and number of turns\\ \midrule
		$L_d$, $L_\lambda$ & \makecell{lists of edges $e$ descending ranked by their \\demand and connectivity increment} \\
		\bottomrule
 	\end{tabular}}
 \vspace{-1em}
\end{table}

\begin{definition1}(\textbf{Transit Network Connectivity})\label{def:conn}
Given a transit network graph $G_r$, the natural  connectivity of $G_r$ is: 
	\vspace{-.8em}
	\begin{equation}
	\label{equ:conn}
	\lambda(G_r) = \ln(\frac{1}{n}\sum_{j=1}^{n}e^{\lambda_j})
	\end{equation}
	Accordingly, $\lambda_1\ge \lambda_2 \ge \cdots \ge \lambda_n$ denotes a non-ascending order of the eigenvalues of $G_r$'s adjacency matrix $\bm{A}$,\footnote{After exponentiation and re-normalization, natural connectivity can be seen as an extension of the \textit{Estrada index} \cite{Estrada2000} ($EE=\sum_{j=1}^{n}e^{\lambda_j}$) which is also widely used in chemistry for measuring the structure of protein.}
	and $n=|V_r|$ is the number of vertexes in $G_r$.
\end{definition1}

\subsubsection{Commuting Demand Measure}
We choose to extend an edge-based trajectory similarity measure in \cite{Wangaclustering}, which is the state-of-the-art for network-constrained trajectories.
With this measure we define the \textit{commuting demand} for a bus route as:

\begin{definition1}(\textbf{Commuting Demand}) Given a set of trajectories $D=\{{T}_1,{T}_2,\cdots,{T}_m\}$ in a road network, and a new bus route $\mu$, we denote the commuting demand that can be met by $\mu$ as:
	\begin{equation}
	\label{equ:dis}
	O_d(\mu)={{\sum_{T_i\in D}{|T_i \cap \mu|}}}
	\end{equation}
where $T_i \cap \mu$ denotes the common edges that $T_i$ and $\mu$ share.
\end{definition1}

\subsubsection{Planning New Route}
With the definitions of transit connectivity and commuting demand measures, 
the objective of \fairbus is to find a new path to optimize both, which is formally defined as:
\begin{definition1}(\textbf{CT-Bus})
\label{def:fairbus}
Given a set of trajectories $D=\{{T}_1,{T}_2,\cdots,{T}_m\}$ in a transit network $G_r$,
\fairbus aims to find a path $\mu$ as 
a new bus route with at most $k$ edges,\footnote{{In our following problem setting, we have multiple real-world constraints which may prune all the candidate paths with $k$ edges, such as turns and circle-free. As a result, we may not get a feasible route with exactly $k$ edges. Thus, we set a more flexible rule on the number of edges here.}} to maximize the following {weighted objective value}:
\begin{equation}
\label{equ:objective}
O = \argmax_{\mu\in G_r^{'} }\left( w\cdot\frac{O_d(\mu)}{{d}_{max}}+ (1-w)\cdot\frac{ O_\lambda(\mu) }{{\lambda}_{max}}\right) 
\end{equation}
where $\mu$ should be a circle-free path both in $G_r^{'}$ and $G$,\footnote{{It means all stops except the departure station should be crossed by the new planned path only once, so {one-way loop is allowed}.}} and $G_r^{'} =\{V_r, E_r^{'}\}$ is the new transit network enriched by new edges in $\mu$, and $O_\lambda(\mu) = \lambda(G_r^{'})-\lambda(G_r)$ is the connectivity increment with $\mu$. 
\end{definition1}

In real bus route planning, two bus stops should not be too far \cite{far} and turn-around should not be frequent \cite{fre}.
Based on the statistics on NYC, two neighbor stops usually have a threshold $\tau$ on the straight line distance and number of turns $tn(\mu)$ \cite{Wang2018c} (see Figure~17 of \cite{Wang2018c}).
We select a fixed constant $\tau = 0.5km$ for \fairbus and 
set a threshold ${Tn}$ on the number of turns of $\mu$, i.e., $tn(\mu)\le {Tn}$.

{Combining multiple objectives into a single weighted objective value is the most straightforward way and has been widely applied in transit route planning \cite{Sharma2009,Bast2015,Weng2020,Fan2006}.}\footnote{{We adopt a linear combination to weigh these two objectives, where a configurable parameter $w$ can meet various planning requirements \cite{Geisberger2010,Batz2012}.
The capacity of bus is not considered here as it is usually considered in a subsequent task after the planned route is determined, i.e. optimal timetable scheduling \cite{Mo2021}.}}
Here, two constants $\lambda_{max}$ and $d_{max}$ are used to normalize two dimensions into the same scale (choices will be discussed in Equation~\ref{equ:newb} for experimental setting).
There are two parameters that are independent of the data and can be set by the users: 1)
$w\in [0, 1]$ is a constant value to balance two objectives, {we set $w=0.5$ by default}; 2) $k$ is the number of edges. 

\section{Our Methods}
\label{sec:fairbus}


\subsection{NP-Hardness of {CT-Bus}}
\label{key}
\fbox{
	\parbox{0.46\textwidth}{
\begin{lemma1}
	\label{lem:np}
	\textbf{\fairbus is NP-hard.} 
\end{lemma1}}
}


{
\begin{proof}
Considering the extreme case where the transit network $G_r$ is a fully connected graph (complete graph), the natural connectivity of $G_r$ will not change when we add a new bus route into the network. Hence, the objective of \fairbus is degenerated to the case where only the commuting demand is considered in bus route planning. 
To simplify the calculation, we convert the demand objective function (Equation~\ref{equ:dis}) into a function of $f_e$, denoting the number of trajectories that include edge $e$. We then have:
\begin{equation}\label{Eq. transit demand}
\begin{aligned}
\begin{split}
O_d(\mu) & ={\sum_{e\in\mu}\sum_{T_i\in D}{b_{i,e}\cdot|e|}},\ where \ b_{i,e}=\begin{cases} 1, \ \ if\ e\in T_i \\ 0, \ \ otherwise \end{cases} \\ &= {\sum_{e\in\mu}{f_e\cdot|e|}}
\end{split}
\end{aligned}
\end{equation}

As a result of such a conversion, we can reduce \fairbus to a routing problem that aims to maximize $\sum_{e\in\mu}{f_e\cdot|e|}$ when constructing a route $\mu$ with $k$ edges. This is equivalent to the {\textit{$k$ minimum traveling salesman problem (k-TSP)} that has been proved to be NP-hard} \cite{Arora2003,Garg2005,Wangaclustering}, where renormalization can be conducted to convert maximization to minimization. More generally, when both commuting demand and network connectivity are considered, the problem is at least as hard as the $k$-TSP problem, and thus is also NP-hard. 
\end{proof}

}

Solving the \fairbus problem is challenging even in the extreme case where only the commuting demand is considered (Equation~\ref{Eq. transit demand}). An approximation algorithm with bounded error can be achieved only if the edge weight satisfies the triangle inequality \cite{Garg2005}, which unfortunately does not hold for  the edge weight (i.e., $f_e\cdot|e|$) in our case.
Hence, we propose an expansion-based greedy algorithm.

\subsection{Expansion-based Traversal Algorithm}


\myparagraph{\underline{Main Idea}} We employ an  expansion-based graph traversal method \cite{Gunawan2016,Wangaclustering} to solve \fairbus, with new optimizations proposed for acceleration, as shown in Algorithm~\ref{alg:fairbus}. 
We first select edges with high demand in the graph for expansion in the \textbf{\textit{initialization}} stage. In the \textbf{\textit{expansion}} phase, we scan candidate paths by initializing  candidate seeds with all edges (including existing edges in $G_r$ and potential edges with a length less than a threshold $\tau$), and then incrementally add neighbor edges as new candidates. In the \textbf{\textit{verification}} phase, we compute the connectivity of each candidate, and update the result if the candidate has a higher objective value, after passing the \textit{feasibility} and \textit{domination} checking. The above two phases will be conducted iteratively until meeting the termination criterion, i.e., the number of iterations exceeds a predefined threshold or there is no more candidate in the queue (Line~\ref{line:termi}). 



\label{sec:expan}



\begin{algorithm}[h]\small
	\caption{\textbf{ETA}($G$, $G_r$, $L_d$)}
	\KwOut{$\mu$: new path}
	\label{alg:fairbus}
	Priority queue $Q\leftarrow \emptyset$, {domination table} $DT\leftarrow \emptyset$, {initial objective value} $O_{max}\leftarrow 0$, {iteration counter} $it\leftarrow 0$\;
	\tcc{\small Initilize candidate edges}
	\texttt{Initialization}($G$, $G_r$, ${Q}$, $\tau$)\;

	\While{$Q\neq \emptyset$}
	{\tcc{\small Scan every candidate path $cp$ from $Q$}
		$\big(\ub(cp), cp, O(cp), tn(cp), cur \big) \leftarrow Q.poll()$\label{line:pqe}\;
		\If{$\ub(cp) \le {max}$ or $it\ge it_{\svar{max}}$\label{line:termi}}
		{
			break\;
		}
		
		\tcc{\small Expand candidate $cp$ with best neighbors}
		$it\leftarrow it+1$, $max_c\leftarrow 0$\;
		\For{each neighbor edge $e\in L_d$ of $cp$'s two ends\label{line:neighbor}}
		{
			\If{$e \notin cp$}{
				$p \leftarrow cp+e$, compute $O(p)$ by Lanczos method\label{line:ops1}\;
				\If{$O(p)>max_c$}
				{
					$max_c\leftarrow O(p)$,
					update $be$ ($ee$) with $e$\;
				}
			}
		}
		\tcc{\small Update the best path $\mu$ with new $cp$}
		$cp \leftarrow be+cp+ee$, compute $O(cp)$ by Lanczos method\label{line:ops}\;
		\If{$O(cp)>{max}$}
		{
			${max}\leftarrow O(cp)$,
			$\mu \leftarrow cp$\;
		}
		\tcc{\small Insert $cp$ into $Q$ for further expansion}
		\texttt{FurtherExpansion}($cp$, $O(cp)$, ${Q}$)\label{line:ffe}\;
	}
	\Return $\mu$\;
	\vspace{-0.5em}
	\algrule
	\vspace{-0.5em}
	\SetKwFunction{FMain}{Initialization}
	\SetKwProg{Fn}{Function}{:}{}
	\Fn{\FMain{$G$, $G_r$, ${Q}$, $\tau$}}{
		$L_d\leftarrow$ \textsc{CandidateEdges}($G_r$, $\tau$, $G$)\label{line:init}\;
		\For{each edge $e_i$ in $L_d$}{
			Update $\mu$ and $O_{max}$ by $e_i$ and $O(e_i)$ \label{line:oe}\;
			$cur \leftarrow k$, $\ub_d(e_i)\leftarrow \sum_{i = 1}^k L_d(i)$\;
			\If{$i>k$}
			{
				$cur \leftarrow k-1$\; $\ub_d(e_i)\leftarrow \ub_d(e_i)-(L_d(k)-L_d[e_i])$\;
			}
			$\ub(e_i) \leftarrow w\cdot \frac{\ub_d(e_i)}{d_{max}}+(1-w)\cdot \frac{\ub_\lambda(e_i)}{\lambda_{max}}$\label{line:ub}\;
			$Q.push\big(\ub(e_i), e_i, O(e_i), 0, cur\big)$\;
		}\label{line:init2}
	}
	\vspace{-0.5em}
	\algrule
	\vspace{-0.5em}
	\SetKwFunction{FMain}{FurtherExpansion}
	\SetKwProg{Fn}{Function}{:}{}
	\Fn{\FMain{$cp$, $O(cp)$, ${Q}$}}{
		\If{ $tn(cp)<\var{Tn}$ \& $\ub(cp)>O_{max}$ \& $len(cp)<k$\label{line:start}}{
			Update $\big(\ub_d(cp)$, $tn(cp)$, $cur\big)$ by Algorithm~\ref{alg:bound}\label{line:lbf}\;
			$\ub(cp) \leftarrow w\cdot \frac{\ub_d(cp)}{d_{max}}+(1-w)\cdot \frac{\ub_\lambda(cp)}{\lambda_{max}}$\label{line:ub2}\;
			\If{$O(cp)>DT(cp.be, cp.ee)$}
			{
				$DT(cp.be, cp.ee) \leftarrow O(cp)$\label{line:buffer}\;
				$Q.push\big(\ub(cp), cp, O(cp), tn(cp), cur\big)$\label{line:pq}\;
			}
		}\label{line:end}
	}
\end{algorithm}

\subsubsection{Initialization}
In the Function \texttt{Initialization} (details in Line~\ref{line:init} to \ref{line:init2}), we select all the potential edges with a distance within $\tau$ as the seeding paths for expansion, and insert them to a priority queue $Q$ for further expansion. 
To generate the candidate edges, we find all the neighboring stops that are within a distance $\tau$ of a stop and add the pair into the candidate edge list $L_d$. 
The edges in $L_d$ are sorted in descending order based on their  demand (i.e., $f_e\cdot|e|$), and we use $L_d(i)$ to denote the demand of the $i$-th edge in $L_d$. 
To compute the demand of each new edge, we conduct a shortest path search to connect its two bus stops and aggregate the demands of all the road network edges it crosses. 

\subsubsection{Expansion from Two Ends}
In Line~\ref{line:neighbor}, the expansion is conducted by adding new edges, where we have two options:

\myparagraph{1) All Neighbors} 
Enqueueing the candidate path appended with each new neighbor edge enables us to scan potential candidates fully.
Both depth-first and breadth-first can be used to fully scan all the candidates, and here we employ a breadth-first search based on a priority queue. Appending with each new neighbor will result in a large queue and thereby the algorithm is hard to terminate.
We call the method with this option as \textbf{ETA-AN}. 

\myparagraph{2) Best Neighbor}
To solve the convergence issue, we propose to choose only the best neighbor edge as shown in Algorithm~\ref{alg:fairbus}, i.e., after selecting the optimal neighbors: {the \textbf{b}eginning \textbf{e}dge $be$ and the \textbf{e}nding \textbf{e}dge $ee$} with the highest increment, we let $cp \leftarrow be+cp+ee$ (Line~\ref{line:ops}).
The queue will then remain a proper size without growing, and even become smaller after \textit{feasibility checking}. 

Before inserting the candidate $cp$ into $Q$ for further expansion in the function at Line~\ref{line:ffe} (details in Line~\ref{line:start} to \ref{line:end}), we compute the objective, update the optimal path $\mu$ in Line~\ref{line:ops}, and then estimate its upper bound $\ub(cp)$ with the current best result's score $O_{max}$.
If $\ub(cp)>O_{max}$, this candidate is inserted; otherwise, it is discarded.
To distinguish all candidates in the queue, each will be attached with an upper bound on its objective value (Line~\ref{line:pq}).


\subsubsection{Verification}
From Line~\ref{line:ops}, we compute the objective value of each candidate path $cp$ and check whether it can replace the current best result $\mu$.
Before inserting $cp$ into the queue in Line~\ref{line:pq} for future expansion, we conduct the following checking:

\myparagraph{Feasibility Checking}
Circle-free is a basic criterion in planning bus routes, and every edge can be crossed once in a bus route.
Turn-checking will check how many turns the candidate path already has and will discard it if the number exceeds a threshold $\var{Tn}$. 

\myparagraph{Domination Checking}
In Line~\ref{line:buffer}, all candidates will go to a {\textbf{d}omination \textbf{t}able $DT$} composed of the checked paths, to compare the objective value and conduct the domination checking.
This step can avoid repetitive expansion on the path that shares the same beginning edge $be$ and ending edge $ee$ but with a smaller objective value.
We call the method without this optimization as \textbf{ETA-DT}.

{\myparagraph{\underline{\textit{Running Example}}} Recall Figure~\ref{fig:example} as a toy example, our algorithm first takes a new edge $e$ (dotted line) as a candidate, and starts an expansion with neighbor edge as shown in the red line. To compute the objective value $O$, we will estimate the connectivity of the updated transit network with $e$, and the demand increment $12+7$.}

\subsection{Bottlenecks to Make ETA Work Efficiently}
{The above expansion-based algorithm applies a common methodology in solving route planning problems \cite{Gunawan2016}.
However, there are two main efficiency bottlenecks. 


\myparagraph{Bottleneck 1}There would be intensive connectivity computations (Line~\ref{line:ops1} and \ref{line:ops}), where a single operation will cost minutes (see Column 2 of Table~\ref{tab:fastconn}), and \fairbus always needs thousands of iterations to terminate according to our experiments.\footnote{{Hence, the whole algorithm’s complexity can be denoted as the product of the number of iterations and the complexity of connectivity estimation (see Lemma~\ref{lem:lanczos_approx}).}}

\myparagraph{Bottleneck 2} To differentiate candidates in $Q$, we estimate their \textit{upper bounds} in lines~\ref{line:ub} and \ref{line:ub2} to predict the best case, and choose the one with the highest upper bound to conduct expansion. The bottleneck here is how can we quickly get tight upper bounds $\ub$.}

{To overcome Bottleneck 1, 
	we convert the connectivity computation to fast trace computation using the \textit{Lanczos method} \cite{Ubaru2017,Musco2018}, combined with Hutchinson's stochastic trace estimator \cite{Hutchinson1990}. We then derive tight upper bounds on the objective values based on the estimated connectivity in Section~\ref{sec:computeconn}.
 	To overcome Bottleneck 2, we pre-compute the connectivity increment for every edge in Section~\ref{sec:expopti}, that can further boost the performance of \textbf{ETA}.}


\section{Fast Connectivity and Bound Estimation}
\label{sec:computeconn}
{In this section, we first show how to efficiently calculate the natural connectivity of a transit network by estimating the adjacency matrix's trace (Equation~\ref{eq:trace}), and prove that its approximation error can be bounded within 1\% (Lemma~\ref{lem:lanczos_approx}). Subsequently, we derive two upper bounds on the connectivity of a new network enhanced with a path $\mu$ (Lemma~\ref{lem:general_edge_additions} and \ref{lem:path_edge_additions}). Finally, we introduce a fast way to incrementally update the upper bound based on the previous bound of a path, without recalculating it from scratch (Algorithm~\ref{alg:bound}).}

\subsection{Lanczos-based Connectivity Estimation}
Given an updated graph $G_r$, estimating its connectivity $\lambda(G_r)$ efficiently and precisely is crucial to answering \fairbus.
We solve this problem by converting it to a \textit{matrix trace estimation problem}, which can be rapidly approximated by combining \emph{the Lanczos method} \cite{Ubaru2017,Musco2018,Dong2019,Ubaru2018,Beckermann2018}2 with \emph{Hutchinson's stochastic trace estimator} \cite{Hutchinson1990,avron2011,Dharangutte2021,Meyer2021}. These techniques are often used together in the applied mathematics literature and turn out to be a winning combination for accurately estimating natural connectivity in this paper. 
Specifically, we start with the observation that:
	\begin{equation}
	\label{eq:trace}
	\lambda(G_r) = \ln(\frac{1}{n}\sum_{j=1}^{n}e^{\lambda_j}) = \ln(\frac{1}{n}\tr(e^{\bm{A}}))
	\end{equation}
	
\noindent where $\tr$ is the matrix trace and $e^{\bm{A}} \in \mathbb{R}^{n\times n}$ is the standard matrix exponential of the adjacency matrix $\bm{A}$. So our task reduces to approximating $\tr(e^{\bm{A}})$. One approach to do so, which was taken in prior work \cite{Chan2014,Chen2018d}, is to compute only the largest eigenvalues of $\bm{A}$ using an iterative Krlyov subspace method (like the Lanczos method) and to estimate Equation~\ref{eq:trace} using a truncated sum. However, for transit networks, which are typically planar or nearly planar, the eigenvalues of $\bm{A}$ decay very slowly, so many eigenvalues are needed to accurately approximate $\lambda(G_r)$. 

Fortunately, the Lanczos method can be used in a far more economical way. \citet{Hutchinson1990} made the powerful observation that, for any $\bm{M} \in \mathbb{R}^{n\times n}$,
	\begin{equation}
        \mathbb{E}(\bm{v}^T\tr(\bm{M})\bm{v}) = \tr(\bm{M})
	\end{equation}
when $\bm{v}$ is a vector with unit variance random Gaussian entries. Accordingly, if we draw $s$ random Gaussian vectors $\bm{v}_1, \ldots, \bm{v}_s$, we can estimate $\tr(\bm{M})$ by:
\begin{equation}
    \gamma = \frac{1}{s}\sum_{i=1}^s \bm{v}_i^T\tr(\bm{M})\bm{v}_i
\end{equation}
It is possible to prove that, if $\bm{M}$ is positive semi-definite and $s = O(\log(1/\delta)/\epsilon^2)$, then with probability $(1-\delta)$, $\gamma$ is within a multiplicative $(1\pm \epsilon)$ of $\tr(\bm{M})$ \cite{Roosta-Khorasani2013}.
Since $e^{\bm{A}}$ is always positive semi-definite, this bound immediately applies to our problem. 

What's more, the required computation of $\bm{v}^T\tr(e^{\bm{A}})\bm{v}$ can be accelerated by \emph{iteratively} approximating $e^{\bm{A}}\bm{v}$ using the Lanczos method for the matrix exponential. Each iteration of this method requires a matrix vector multiply with $\bm{A}$, which takes just $O(m)$ time, where $m$ is the number of edges in $G_r$. 
To bound the number of iterations of Lanczos needed for an accurate approximation, we  state a corollary\footnote{This corollary follows from a simple algebraic manipulation of Theorem 15 in \cite{Musco2018} (see also \cite{OrecchiaSachdevaVishnoi:2012}), combined with the fact that $\tr(e^{\bm{A}}) \geq e^{\|\bm{A}\|_2}$.} of Theorem 15 in \citet{Musco2018}:

\noindent\fbox{
	\parbox{0.46\textwidth}{
\begin{lemma1}[\textbf{Lanczos Approximation Bound}]
\label{lem:lanczos_approx}
After $t = O\left(\|\bm{A}\|_2 + \log(1/\epsilon)\right)$ iterations, 
the Lanczos method returns an approximation 
$\bm{s}$ to $e^{\bm{A}}\bm{v}$ satisfying:
\begin{equation*}
    \|\bm{s} - e^{\bm{A}}\bm{v}\|_2 \leq \epsilon \tr(e^{\bm{A}})\|\bm{v}\|_2
\end{equation*}
\end{lemma1}}}
\vspace{1em}

Above $\|\bm{A}\|_2$ denotes the spectral norm of $\bm{A}$. Even when $\bm{A}$ is large, this is typically very small for transit networks (and planar graph adjacencies more generally). For example, for the Chicago and NYC transit networks analyzed in this paper, $\|\bm{A}\|_2$ equals $5.46$ and $4.79$, respectively. Accordingly the number of iterations required to accurately approximate $e^{\bm{A}}\bm{v}$ essentially depends just logarithmically on the desired accuracy $\epsilon$.

Combined with the stated bound on the number of samples needed for Hutchinson's estimator, and using that for a scaled Gaussian random vector $\|\bm{v}\|_2 = O(\sqrt{n})$ with high probability, we conclude that $\tr(e^{\bm{A}})$ can be estimated to multiplicative $(1\pm\epsilon)$ error with $O(\log(1/\delta)/\epsilon^2)$ approximate computations of $\bm{v}^T e^{\bm{A}}\bm{v}$, each of which takes just $O\left(\|\bm{A}\|_2 + \log(n/\epsilon)\right)$ iterations. This translates to an additive $\pm \epsilon \ln(\tr(e^{\bm{A}}))$ approximation to $\lambda(G_r)$.

Experimentally, we confirm the low complexity of the Lanczos + Hutchinsons method. In this paper, we use a default setting of $s = 50$ repetitions of Hutchinson's estimator, each computed using $t =10$ iterations. We typically obtain an approximation to $\lambda(G_r)$ accurate to with  \underline{$\bm{1\%}$  \textbf{error}}. 
Table~\ref{tab:fastconn} shows the time comparing with eigenvalues-based full computation.


\begin{table}
	\ra{1}
	\caption{Running time of connectivity \& bound estimation.}
	\vspace{-1em}
	\label{tab:fastconn}
	\scalebox{1}{\begin{tabular}{cccccc}
		\toprule
		\textbf{City} &   \textbf{\makecell{Eigen\\NumPy}} & \textbf{\makecell{Lanczos\\NumPy}} &\textbf{\makecell{Lanczos\\Matlab}} & \textbf{\makecell{General \\bound}}& \textbf{\makecell{Path \\bound}} \\
		\midrule
		Chicago &  28.65s &0.610s  & 0.035s& 0.102s  & 0.049s \\
		NYC &  225.03s& 2.412s  & 0.094s& 0.204s&0.099s  \\
		\bottomrule
	\end{tabular}}
\end{table}

\subsection{Connectivity Upper Bound Estimation}
\label{sec:norbound}
\citet{DeLaPena2007} estimated bounds for the Estrada index, which can be converted to the maximum natural connectivity with $k$ arbitrary edges:
\vspace{-0.5em}
\begin{align*}
\lambda(G_r') \leq \ln(1+\frac{e^{\sqrt{2(|E_r|+k)}}-1}{|V_r|})
\end{align*}
However, we found that this bound is much bigger than the real connectivity, and it is too loose to be used as a normalization value (see a comparison in Table~\ref{tab:bound} where $k=15$).

\begin{table}
	\ra{1}
	\caption{Tightness comparison of connectivity upper bound.}
	\vspace{-1em}
	\label{tab:bound}
	\scalebox{1}{\begin{tabular}{ccccc}
			\toprule
			\textbf{City} &   \textbf{\makecell{Estrada \\ bound \cite{DeLaPena2007}}} & \textbf{\makecell{General\\ bound}} &\textbf{\makecell{Bound\\path}} & \textbf{\makecell{Increment\\bound}}  \\
			\midrule
			Chicago & 104.205 & 1.576 & 0.167 &0.034 \\
			NYC &  156.459 & 0.655 &0.067 &0.010\\
			\bottomrule
	\end{tabular}}
	\vspace{-1em}
\end{table}

We further propose a tighter upper bound on the connectivity after adding $k$ edges, which will depend on the connectivity of the original graph, and top-$k$ eigenvalues of the original graph adjacency matrix $\bm{A}$. This can be computed quickly using a Lanczos method that we have mentioned.
Further, one of the advantages of the expression is that it gets much tighter if the $k$ edges added in \fairbus form a path.
This bound will also help estimate the upper bound of a candidate to be filled with less than $k$ edges, and solve existing or new network optimization problems \cite{Chan2014,Chen2018d} in future.

\noindent\fbox{
\parbox{0.46\textwidth}{
\begin{lemma1}[{\textbf{General Upper Bound}}]\label{lem:general_edge_additions}
If $G_r'$ is obtained by adding $k$ arbitrary unweighted edges to $G_r$, the natural connectivity satisfies:
\begin{align*}\small
    \lambda(G_r') \leq \ln\left(e^{\lambda(G_r)} - \sum_{i=1}^{2k} e^{\lambda_i}  + \frac{e^{\lambda_1}}{n}\left[e^{\sqrt{2k}} + 2k-1\right] \right)
\end{align*}
where $\lambda_1 \geq \ldots, \lambda_{2k}$ are the $2k$ algebraically largest eigenvalues of $G_r$'s adjacency matrix. 
\end{lemma1}}}
\vspace{1em}

{The proof of this lemma can be found in the appendix of our technical report \cite{Wang2021Bus}.}

To obtain a tighter upper bound when edges are specifically added into a path, we rely on Fan's powerful generalization of Weyl's inequality \cite{Fan:1949,Bhatia:2001}. This bound tells us that, if $\lambda_1 \geq \ldots \geq \lambda_n$, $\lambda_1' \geq \ldots \geq \lambda_n'$, and $\sigma_1 \geq \ldots \geq \sigma_n$ are any non-increasing ordering of the eigenvalues of $\bm{A}$, $\bm{A}'$, and $\bm{K}$, respectively, then:
\begin{align}
\label{eq:ky_fan}
&\text{For all $q \in 1,\ldots,n$} & \sum_{i=1}^q \lambda_i' \leq \sum_{i=1}^q \lambda_i + \sum_{i=1}^q \sigma_i
\end{align}

\noindent\fbox{
	\parbox{0.46\textwidth}{
\begin{lemma1}[{\textbf{Upper Bound for Paths}}]\label{lem:path_edge_additions}
If $G_r'$ is obtained by adding a $k$ edge simple path to $G_r$, then the natural connectivity satisfies:
\begin{align*}\small
    \lambda(G_r') \leq  \ln\left(e^{\lambda(G_r)} + \frac{1}{n}\sum_{i=1}^{\lfloor\frac{k+1}{2}\rfloor} (e^{\sigma_i} - 1)e^{\lambda_i} \right) 
\end{align*}
where $\sigma_i = 2\cos\left(\frac{i\pi}{k+2}\right)$ is the $i^\text{th}$ eigenvalue of the path graph adjacency matrix. 
\end{lemma1}}}
\vspace{1em}

{The proof of this lemma can be found in the appendix of our technical report \cite{Wang2021Bus}.}

\vspace{0.5em}
Note that computing the bound of Lemma \ref{lem:path_edge_additions} requires computing the top $O(k)$ eigenvalues of $\bm{A}$, which can be done to high accuracy in roughly $O(|E|\cdot k\cdot \log n)$ time for a graph with $|E|$ edges \cite{MuscoMusco:2015}.
In Algorithm~\ref{alg:fairbus}, $\ub_\lambda(cp)$ and $\ub_\lambda(e_i)$  are both computed based on  Lemma~\ref{lem:path_edge_additions}.

\subsection{Incremental Demand Bound Estimation}
\label{sec:queue}
We set the initial upper bound of the demand as the sum of the top-$k$ edge demands in $L_d$: 
\begin{align*}
O_d(\mu) \leq \sum_{i = 1}^k L_d(i)
\end{align*}

When a new edge is added, the path is updated, and we need to re-estimate this upper bound.
A baseline is to re-scan the whole path and enrich the rest uncovered edges by $k-len(cp)$ top edges in $L_d$, which should not be in $cp$, where $len(cp)$ denotes number of edges in $cp$, and $E_r(i)$ denotes $i$-th edge in $L_d$. 
\begin{equation}
\ub_d(cp) = \sum_{e\in cp} L_d[e]+\sum_{i = 1 ~\&~E_r(i) \notin cp}^{k-len(cp)} L_d(i)
\end{equation}

However, this is not efficient when it needs to be conducted for every candidate.
Instead of scanning, we propose a dynamic strategy to update $\ub_d(cp)$ while returning the same bound.
When a new edge is added, if its weight $L_d[e]$ is smaller than the $cur$-th top edge's demand $L_d(cur)$, it means we can replace one top edge with the inserted one, then we update $\ub_d(cp)$ by reducing the gap $L_d(cur)-L_d[e]$, the cursor value $cur$ will decrease by one;
otherwise, $\ub_d(cp)$ and $cur$ will not change.
Each cursor $cur$ is initialized as $k$ at the beginning and will be inherited in the iteration with the upper bound.
Based on this incremental method, we can dynamically update the bound without scanning the whole path $cp$ and ranking list $L_d$, which is more efficient and space-saving.
Details are presented in Algorithm~\ref{alg:bound}.

\vspace{-1em}
\begin{algorithm}\small
	\caption{Incremental update on bound \& turn}
	\KwIn{$e$, $\ub_d(cp)$, $tn(cp)$, $cur$}
	\label{alg:bound}
	\If{$L_d(cur)>L_d[e]$}
	{
		$cur \leftarrow cur-1$\;
		$\ub_d(cp) \leftarrow \ub_d(cp)-({L_d(cur)-L_d[e]})$\;
	}
	$angle\leftarrow \textsc{ComputeAngle}(e, cp.end)$\;
	\If{$angle>\frac{\pi}{4}$}
	{
		$tn(cp)\leftarrow tn(cp)+1$\;
	}{
		\If{$angle>\frac{\pi}{2}$}{$tn(cp)\leftarrow \var{Tn}$\;}
	}
	\Return $\big(\ub_d(cp), tn(cp), cur\big)$\; 
\end{algorithm}

\section{Pre-computation for Faster ETA}
\label{sec:expopti}
{Although employing the Lanczos method can boost the efficiency of estimating connectivity and upper bounds in Algorithm~\ref{alg:fairbus}, a single iteration's runtime is still nonnegligible, as shown in Table~\ref{tab:fastconn}.
Hence with thousands of iterations, it is slow to terminate this algorithm (see our experiment results in Column 2 \& 4 of Table~\ref{tab:running}).}
%
Thus, in this section, by reducing the connectivity to a linear aggregation function for fast updates with new edges, pre-computation is conducted on each edge's demand and connectivity, which can accelerate the algorithm drastically by using a greedy strategy.

\subsection{Linear Connectivity and Bound Increment}
\label{sec:ubincre}
Recall Figure~\ref{fig:connectivity} the connectivity of a network keeps decreasing as edges are removed. It inspires us to explore the possibility of having a linear increase in connectivity by adding edges. 
To facilitate our exploration, we define a new concept as below.
\begin{definition1}(\textbf{Edge Connectivity Increment})
	By adding an edge $e$ into the transit network $G_r$, which generates a new graph $G_r^{'}$, the edge connectivity increment is denoted as:
	\begin{equation}
	\Delta(e) = \lambda(G_r^{'}) - \lambda(G_r)
	\end{equation}
\end{definition1}

By adding a new route $\mu$ with multiple edges, the connectivity increment is denoted as $O_\lambda(\mu)$ in Definition~\ref{def:fairbus}.
Discovering the relation between the sum of $\mu$'s each individual edge, $\sum_{e\in \mu}\Delta(e)$ and $O_\lambda(\mu)$, is a fundamental problem in the literature of \textit{maximizing submodular set functions} \cite{Nemhauser1978}.
However, we have an observation: \textit{Natural connectivity is a monotonic but non-submodular function (when adding new edges)}. The analysis is as below.

\begin{figure}
	\centering
	\subfigure[\small Chicago]{\includegraphics[height=2.5cm]{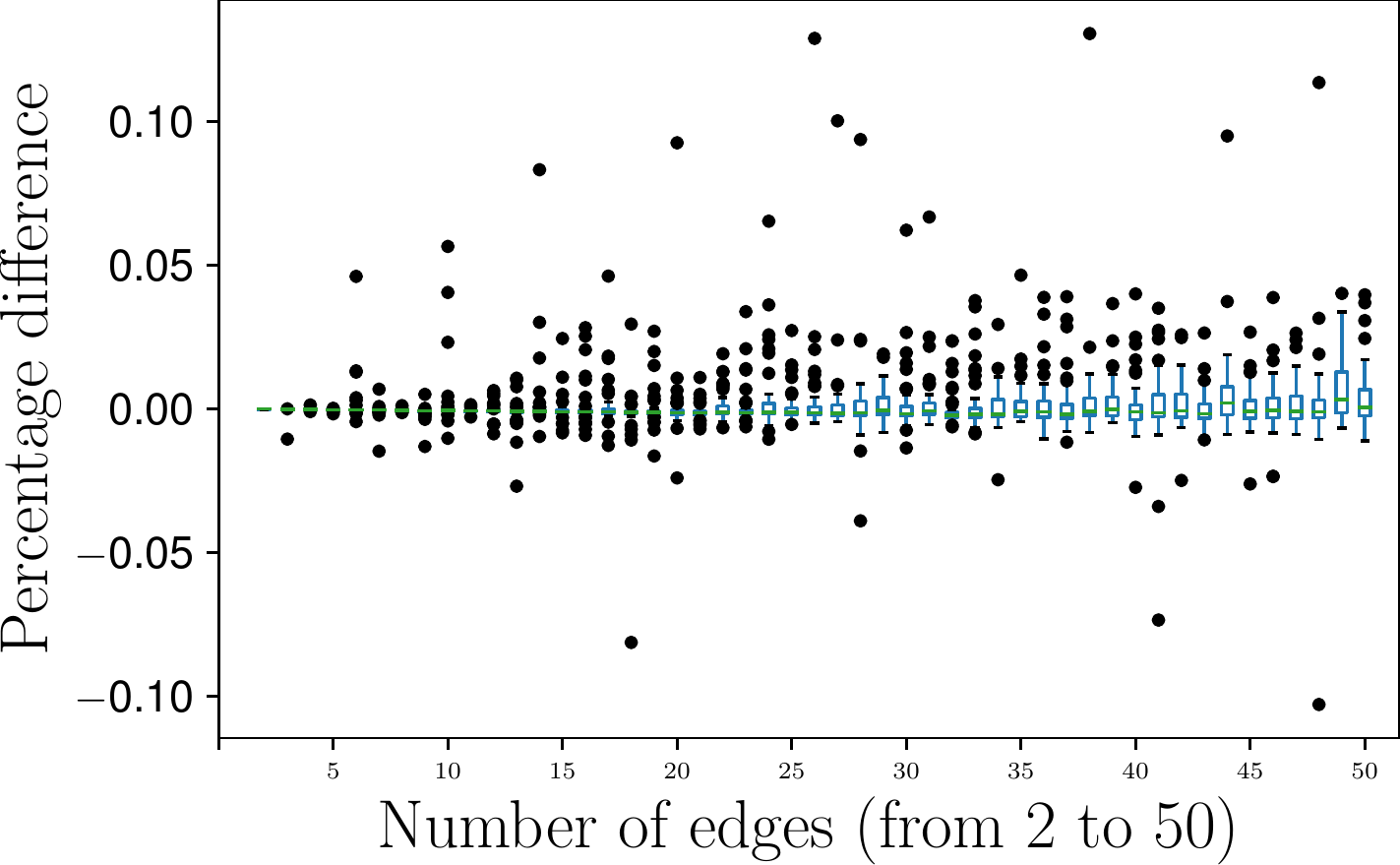}}\hspace{1em}
	\subfigure[\small NYC]{\includegraphics[height=2.5cm]{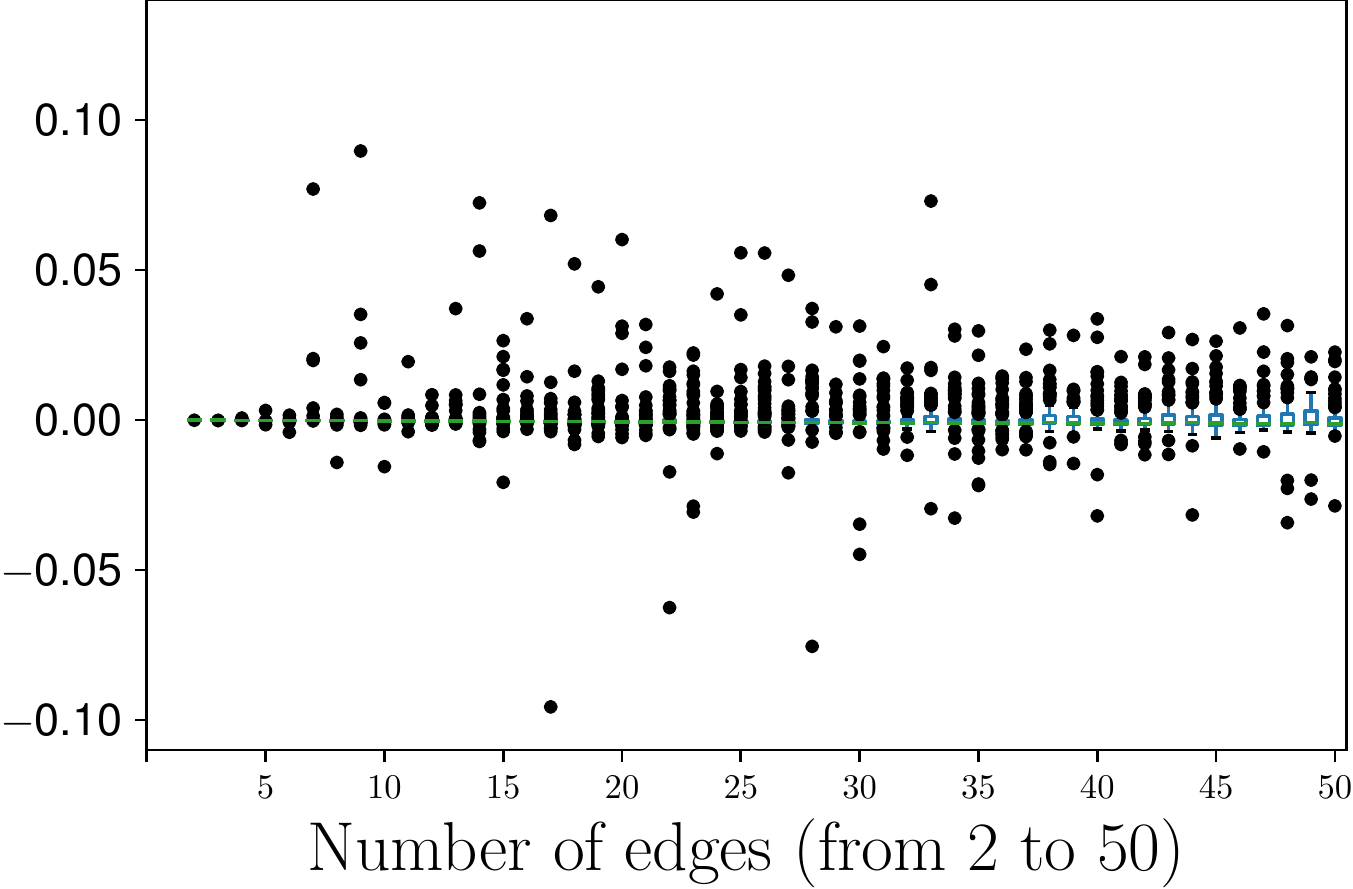}}
	\vspace{-1.5em}
	\caption{Distribution of percentage difference $\theta$ between $O_\lambda(\mu)$ and $\sum_{e\in \mu}\Delta(e)$ with the increasing number of edges.}
	\label{fig:edge-delta}
\end{figure}

Along with the definition, \citet{Wu2010a} have also proved its monotonicity.
To prove it is sub-modular, with the above definition, we just need to verify whether $O_\lambda(\mu)<\sum_{e\in \mu}\Delta(e)$ holds. 
We use counterexamples to verify that natural connectivity is non-submodular. 
As shown in Figure~\ref{fig:edge-delta}, we randomly sample a set of new edges and plot the percentage difference of two connectivity scores: $\theta = \frac{O_\lambda(\mu) - \sum_{e\in \mu}\Delta(e)}{\sum_{e\in \mu}\Delta(e)}$, by increasing the number of selected edges in NYC and Chicago.
The box plot shows that $O_\lambda(\mu)>\sum_{e\in \mu}\Delta(e)$ holds most of the time especially when more edges are included in $\mu$, so natural connectivity is not sub-modular, and there is no guaranteed bound \cite{Nemhauser1978} for a greedy algorithm when answering \fairbus.


Even though natural connectivity is non sub-modular, we still observe that $O_\lambda(\mu)$ is highly close to $\sum_{e\in \mu}\Delta(e)$.
Then, we can use $\sum_{e\in \mu}\Delta(e)$ to estimate the potential connectivity increment to fulfill all the edges in $\mu$, and have: $O_\lambda(\mu) \approx \sum_{e\in \mu}\Delta(e) $.

\subsection{Improvements to Algorithms~\ref{alg:fairbus} and \ref{alg:bound}}
\label{sec:imp}
With the fast Lanczos method, we are able to pre-compute $\Delta(e)$ for all the candidate edges in $L_d$.
Then, we rank them by their connectivity increment $\Delta(e)$ as another descending sorted list $L_\lambda$, i.e.,  $L_\lambda[e]=\Delta(e)$, and $L_\lambda(i)$ returns the $i$-th edge's demand.
Further, the increment upper bound can be estimated using $\ub_\lambda(\mu) = \sum_{i = 1}^k L_\lambda(i)$, similar to $\ub_d$, and its tightness can be observed from the last column of Table~\ref{tab:bound}.
Now, $O_d$ and $O_\lambda$ can be incrementally computed in the same way, and we further combine them into one.



\myparagraph{Integrated Objective Value Increment}
We compute a new objective value composed of the normalized connectivity and demand on each edge, same in Definition~\ref{def:fairbus}, including every existing edge (the connectivity increment is set  as $0$) and every new edge.
Then we rank all the edges by this objective and create a new sorted list $L_e$ in descending order, and convert \fairbus to optimize a single objective instead of two.
With an edge $e$ being added into $cp$, $O(cp)$ will increase by:

\vspace{-0.8em}
\begin{equation}
L_e[e] = w\cdot\frac{L_d[e]}{d_{max}}+(1-w)\cdot\frac{L_\lambda[e]}{\lambda_{max}}
\end{equation}
\vspace{-0.8em}

Next we show how to optimize the algorithms earlier presented in Section~\ref{sec:fairbus}.
First, we replace all the $L_d$ with $L_e$, and $\ub$ with $\ub_d$ in Algorithm~\ref{alg:fairbus} and \ref{alg:bound}, respectively.
Then, we revise Line~\ref{line:oe} as: \textit{Update $\mu$ by neighbor $e_i$ with the highest $L_e(e_i)$}.
Further, we remove Line~\ref{line:ub} and \ref{line:ub2} of Algorithm~\ref{alg:fairbus}.
Lastly, we can simplify line~\ref{line:ops} as: \textit{Increasing $O(cp)$ by $L_e[e]$}, correspondingly. 

\begin{figure}
	\centering
	\hspace{-1em}\subfigure[\small Demand increment]{\includegraphics[height=2.4cm]{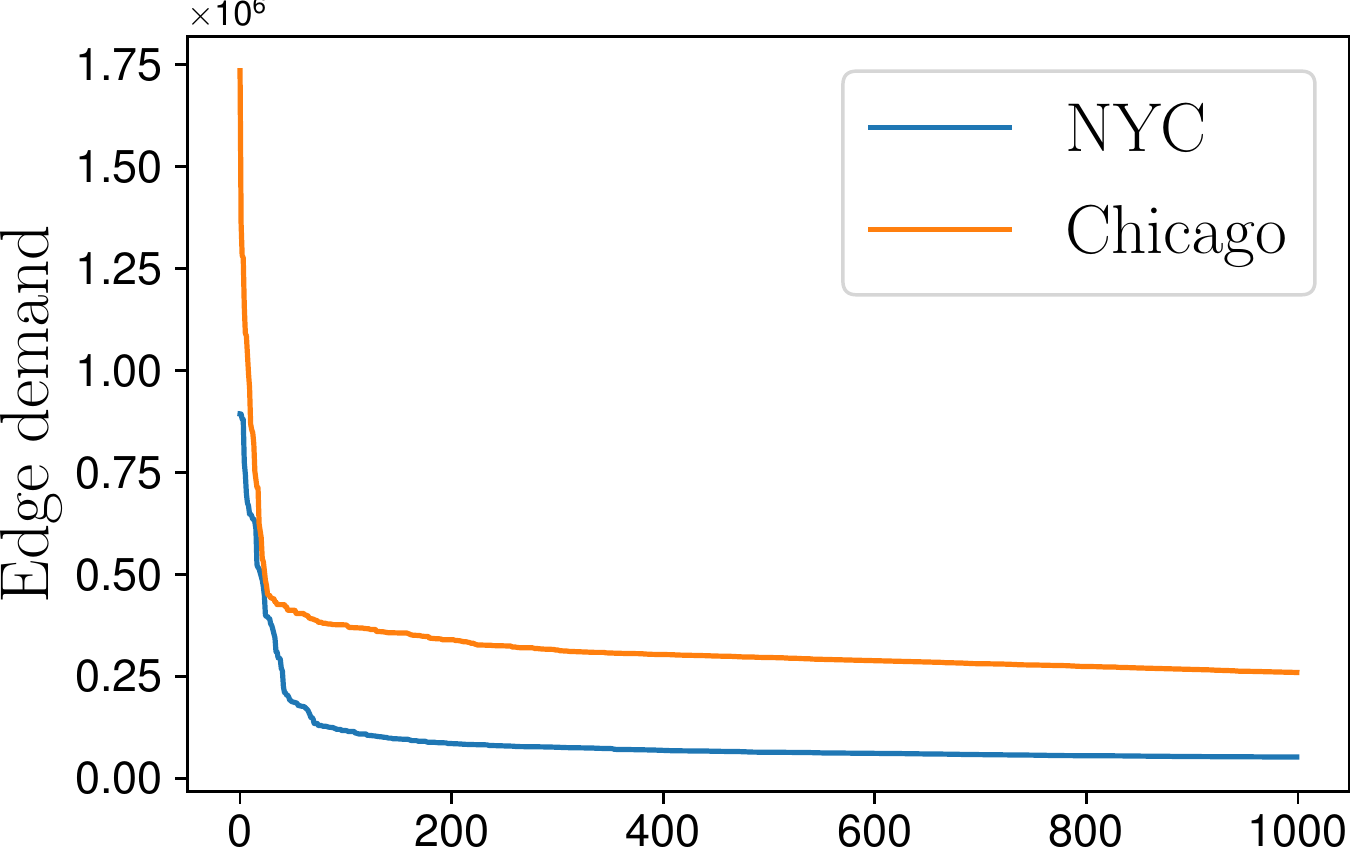}}
	\subfigure[\small Connectivity increment]{\includegraphics[height=2.4cm]{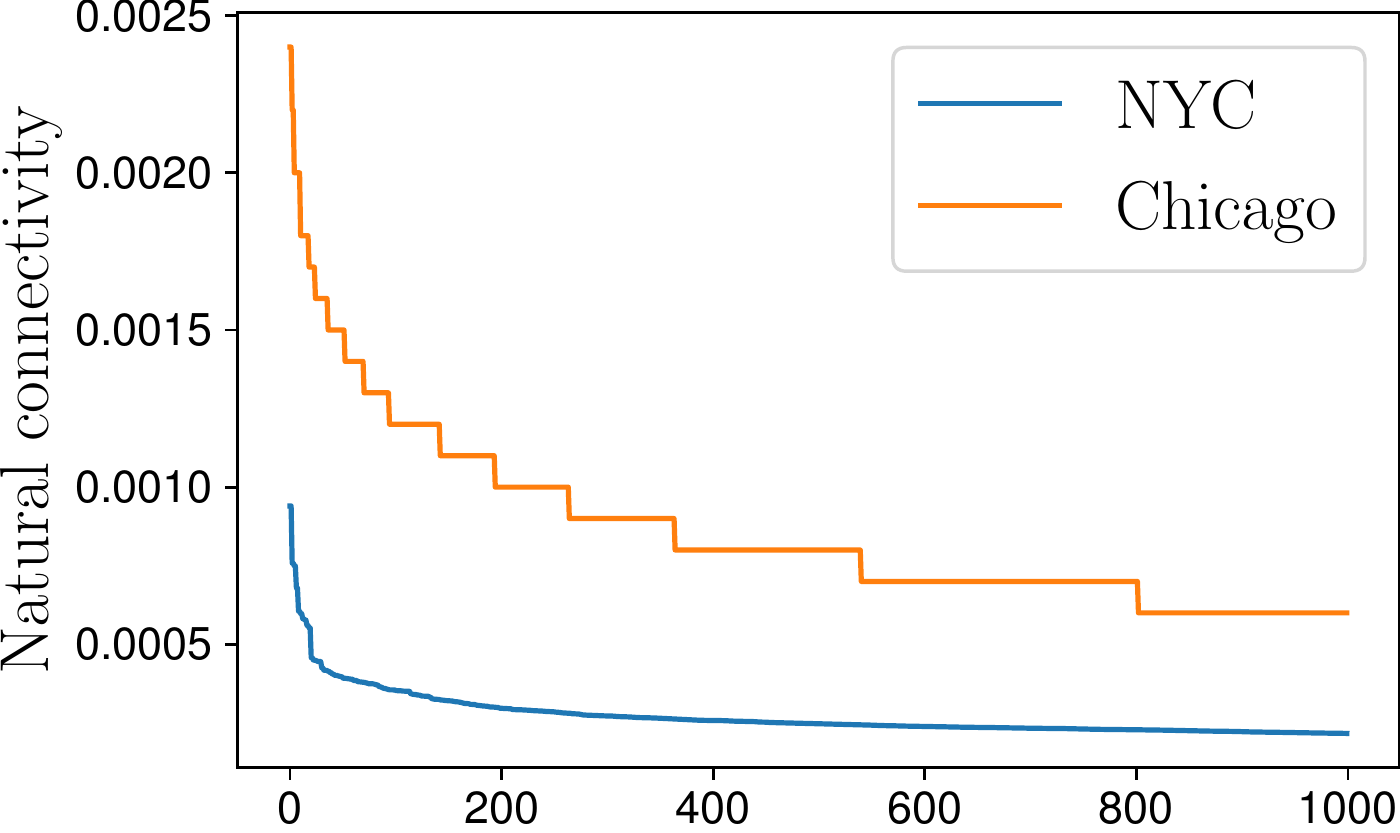}}
	\vspace{-1.5em}
	\caption{Top-$1000$ new edges in terms of increment.}
	\label{fig:topedges}
\end{figure}

\myparagraph{Selective Edges for Seeding}
After our increment computation on edges in $L_d$ using the Lanczos method, we observe that a minority of edges can lead to a large increment on connectivity and demand of the objective functions, as shown in 
Figure~\ref{fig:topedges}.
To reduce the candidate pool size, we choose top-$\var{sn}$ edges in the list $L_e$ as initial seeding paths in Line~\ref{line:init}, where $\var{sn}$ denotes the seeding number.

\subsection{More Discussions}

\myparagraph{Effect of $|D|$ and Pre-processing}
Our method is independent of the number of trajectories $|D|$, since all the trajectories are mapped to the road network and each edge on the road network gets a demand weight.
The pre-processing on mapping and building transit network edge weight is related to the number of new edges, because each edge will invoke one-time shortest path search and one-time connectivity increment estimation (using the Lanczos method).


\myparagraph{Effect of Increasing $\tau$}
According to the above analysis, the complexity of \fairbus is highly related to the number of candidate edges.
A direct parameter to this is the neighbor stop interval distance $\tau$ in Definition~\ref{def:fairbus}; if we increase $\tau$, there will be more candidate edges.
In this paper, we set $\tau$ as a fixed constant ($\tau$=$0.5$km), which is big enough w.r.t. the current statistics in NYC.
However, the number of edges will not sharply increase according to our initial numerical analysis; the running time of pre-computation and ETA will increase linearly and slightly when increasing $\tau$ in a proper range.

\myparagraph{Planning Multiple Routes}
It is worth mentioning that \fairbus can be employed to plan multiple routes as follows -- After planning a new route, we update the graph and its adjacent matrix with the new edges.
Then, we can set all the covered edges' demand value as zero, as our previously-planned new bus routes have covered them.
This step can also be skipped if there is no requirement on whether a new edge should be crossed only once.
At the end, we conduct the new route searching using our algorithm.

\section{Experiments}
\label{sec:exp}
\subsection{Setup}
\subsubsection{Datasets}
\begin{figure*}
	\centering
	\hspace{0.5em}\begin{minipage}{0.6\textwidth}
		\centering
		\subfigure[\small Chicago-Road]{\includegraphics[height=2.5cm]{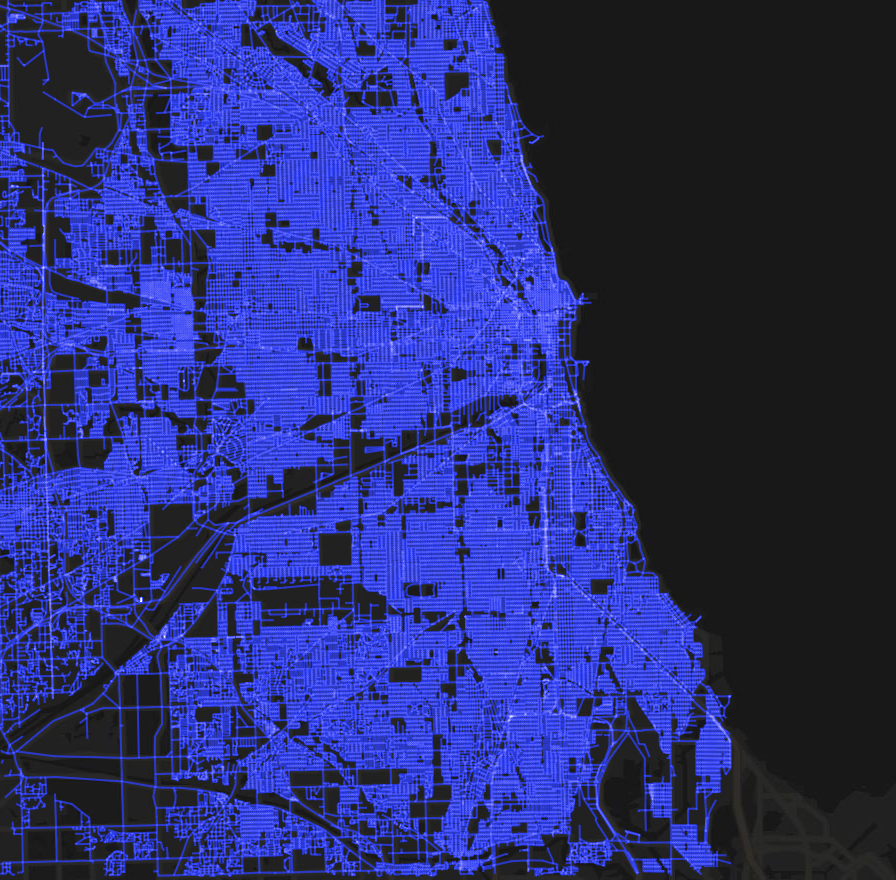}}
		\subfigure[\small Chicago-Transit]{\includegraphics[height=2.5cm]{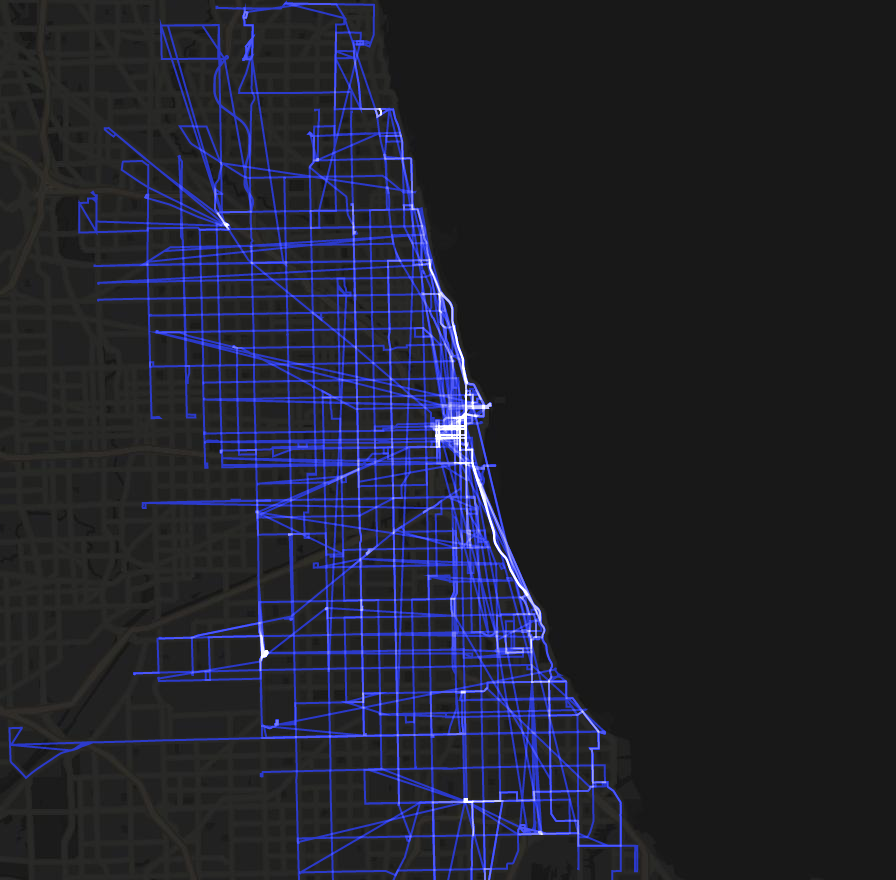}}
		\subfigure[\small NYC-Road]{\includegraphics[height=2.5cm]{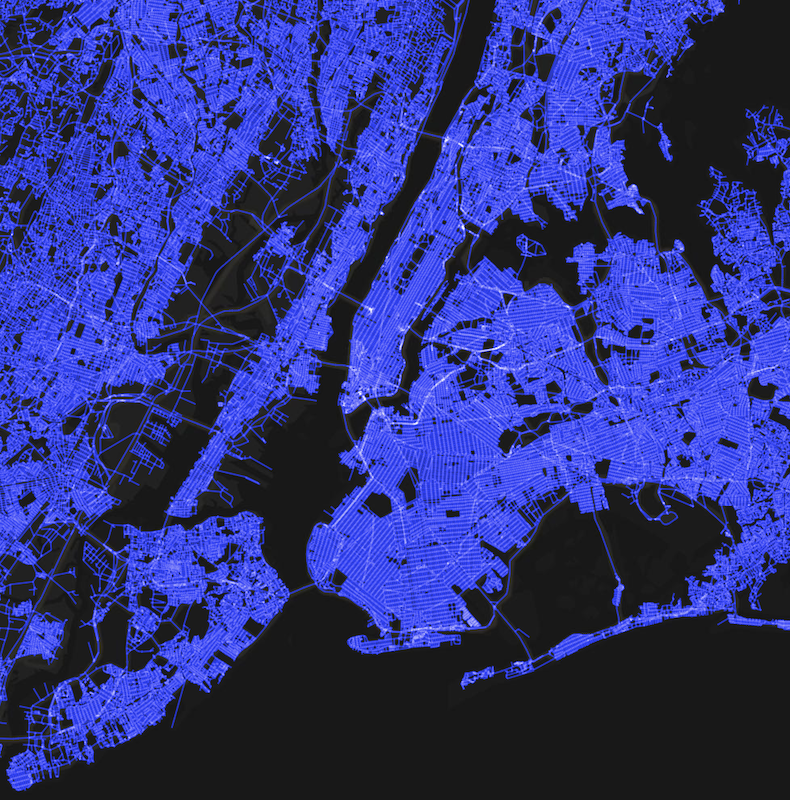}}
		\subfigure[\small NYC-Transit]{\includegraphics[height=2.5cm]{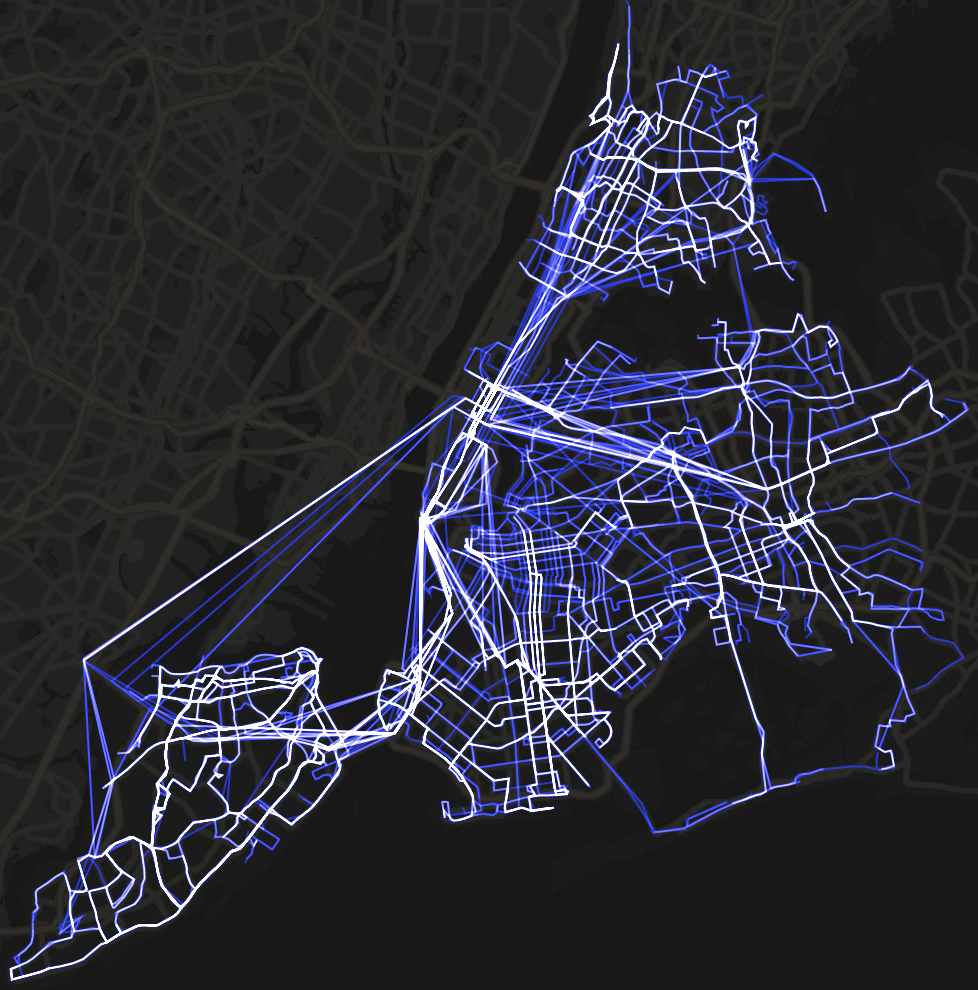}}
		\vspace{-1.5em}
		\caption{An overview of road network and bus network.}
		\vspace{-1em}
		\label{fig:transit}
	\end{minipage}
\hspace{-3.5em}
	\begin{minipage}{0.44\textwidth}
		\centering
		\subfigure[\small Chicago]{\includegraphics[height=2.5cm]{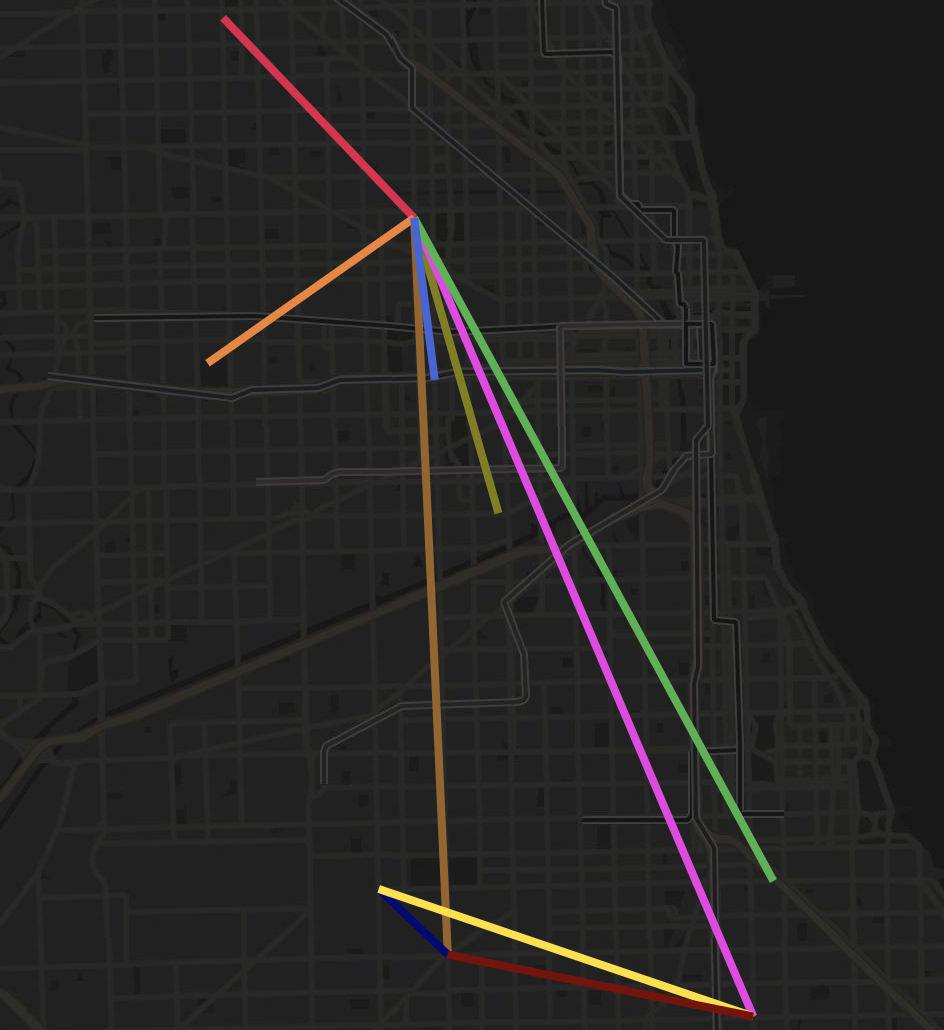}}\hspace{0.7em}
		\subfigure[\small NYC]{\includegraphics[height=2.5cm]{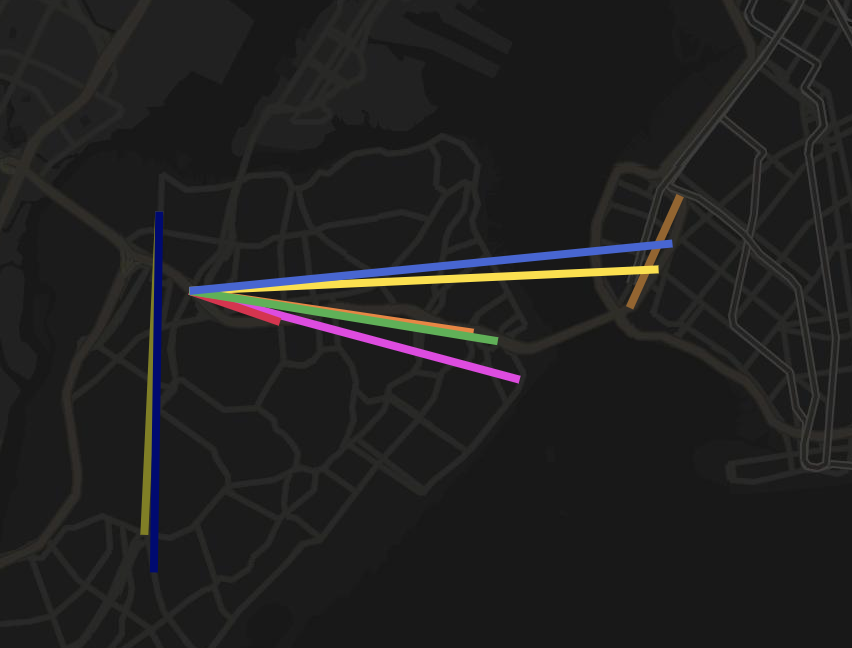}}
		\vspace{-1.5em}
		\caption{Top-$10$ edges of connectivity-first method \cite{Chan2014}.}
		\vspace{-1em}
		\label{fig:baseline-topk}
	\end{minipage}
\end{figure*}

We conduct experiments in two cities, New York City ({NYC}) and Chicago (Chi), {where the road network with travel distance and travel time on each edge} is obtained from DIMACS \cite{DIMACS}, and the transit network is extracted from shapefiles \cite{cta, geonyu}.
Figure~\ref{fig:transit} presents an overview of these four networks.

{Trajectories are obtained from real-world taxi trip records (\cite{chi} for Chicago and \cite{tlc} for NYC) as below: each trip record consists of a pickup and a drop-off location, travel time and travel distance; for each trip we find its shortest path, and if it has a similar travel distance and time (within 5\% error rate) with this trip, we treat it as an approximation of the trip’s real trajectory.
The above way is also used in trajectory-driven site selection problem~\cite{Zhang2018c,Zhang2019a}.
Detailed dataset statistics can be found in Table~\ref{tab:data}.}

\hspace{2.5em}

\subsubsection{Implementation}
We use Python 3 to clean data and search the shortest path between two stops, and MATLAB to compute the connectivity, which is much faster for large matrix computation. 
NetworkX \cite{networkx} is used to store the graph and the shortest path, and
Mapv \cite{mapv} is used to visualize the networks and the returned results in maps.
All experiments are conducted on a laptop with 2.6 GHz 6-Core Intel Core i7 and 32 GB 2400 MHz DDR4 running MacOS Catalina.
Our cleaned datasets, MATLAB code and visualization tools are  available on GitHub for reproducibility \cite{code}.

\subsubsection{Pre-processing}
Table~\ref{tab:precom} shows the number of new edges of our two datasets and their pre-processing time.
Pre-processing time on the candidate new edges includes the time spent on the shortest path search of selective new edges, and their connectivity increment based on the Lanczos method.
Each new edge conducted the shortest path between its two ends, then we put the edge demand by summing up edges in the road network. 
Although this pre-processing is costly, it is called only once for each dataset but will benefit all the algorithms with various parameter settings.

\begin{table}[t]
	\ra{1}
	\caption{Running time of pre-computation on new edges.}
	\vspace{-1em}
	\label{tab:precom}
	\hspace{-1em}
	\centering
	\scalebox{1}{\begin{tabular}{cccc}
			\toprule
			\textbf{Dataset} &  \textbf{\makecell{\#New edges}} & \textbf{Connectivity}  & \textbf{\makecell{Shortest path}} \\
			\midrule
			Chicago & 95,304 & 1857s &15,322s \\
			NYC & 160,790& 7332s  & 33,241s  \\
			\bottomrule
	\end{tabular}}
\end{table}

\begin{table}[t]
	\ra{1}
	\caption{An overview of datasets. {$|R|$: number of bus routes; $len(R)$: average number of stops; $|V|$ and $|E|$: number of vertices and edges in road network; $|V_r|$ and $|E_r|$: number of vertices and edges in bus network; $|D|$: number of trajectories.}}
	\vspace{-1em}
	\label{tab:data}
	\hspace{-1em}
	\centering
	\scalebox{0.88}{\begin{tabular}{cccccccc}
		\toprule
		\textbf{Dataset} & \textbf{\makecell{$|R|$}} & \textbf{$len(R)$} & $|V|$& $|V_r|$ &  $|E|$ & $|E_r|$ & \textbf{$|D|$ }\\
		\midrule
		Chicago & 146 &47& 58,337 & 6171 &89,051 & 6892  & 555,367 \\
		NYC & 463 &30& 264,346 & 12,340 &  365,050 & 13,907& 407,122 \\
		\bottomrule
	\end{tabular}}
	\vspace{-1em}
\end{table}

\subsubsection{Evaluation Metrics}
\begin{figure*}[h]
	\centering
	\begin{minipage}{0.75\textwidth}
		\centering
		\hspace{-0.5em}
		\subfigure[\small Chicago]{\includegraphics[height=4.4cm]{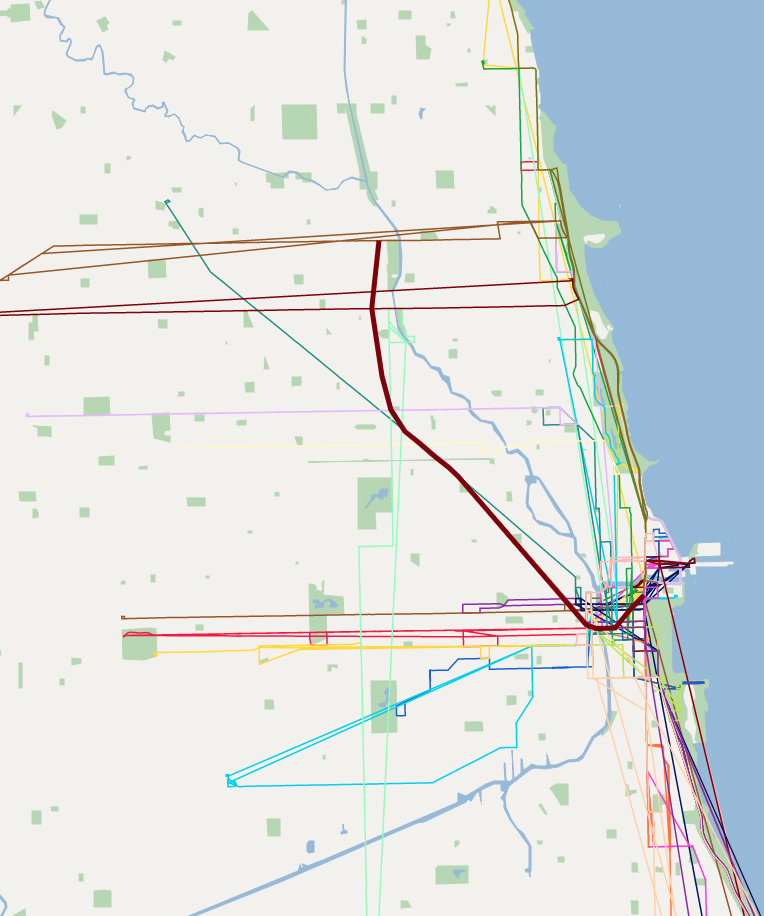}}\hspace{2.8em}
		\subfigure[\small Manhattan]{\includegraphics[height=4.4cm]{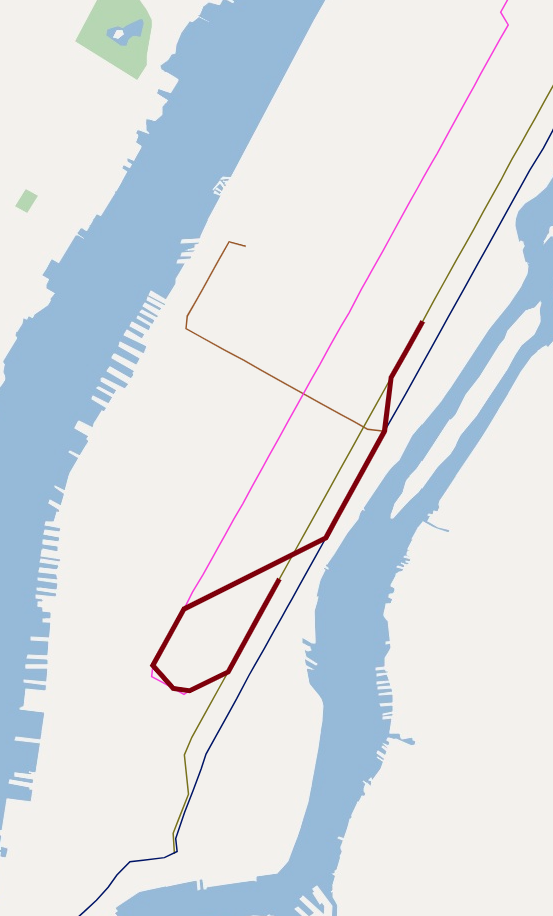}}\hspace{2.8em}
		\subfigure[\small Queens]{\includegraphics[height=4.4cm]{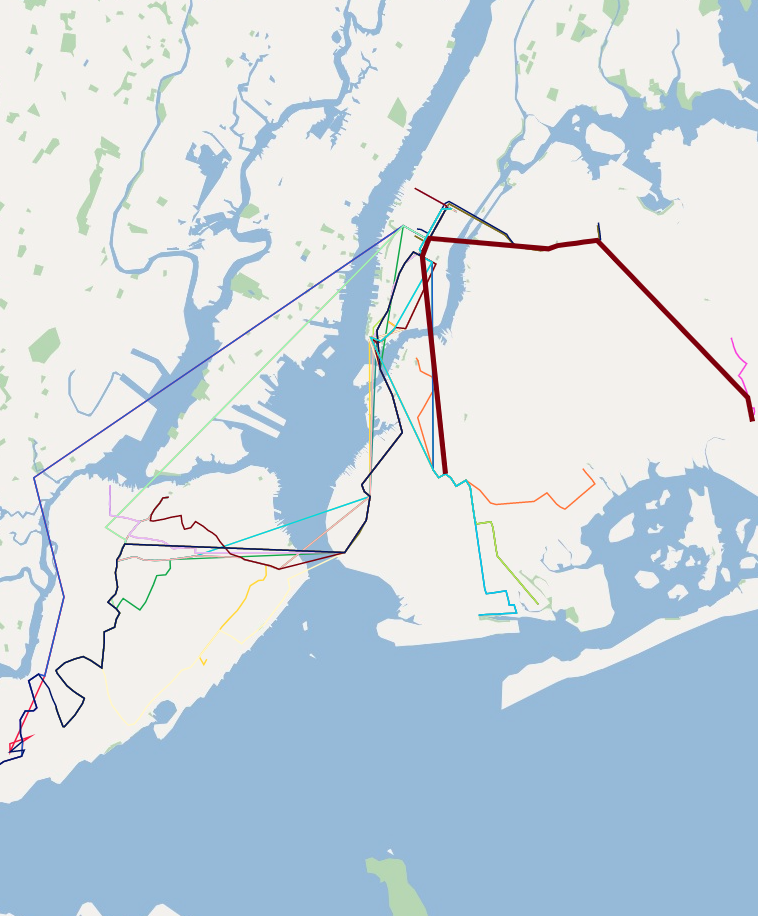}}\\ \vspace{-0.7em}
		\subfigure[\small Brooklyn]{\includegraphics[height=4.4cm]{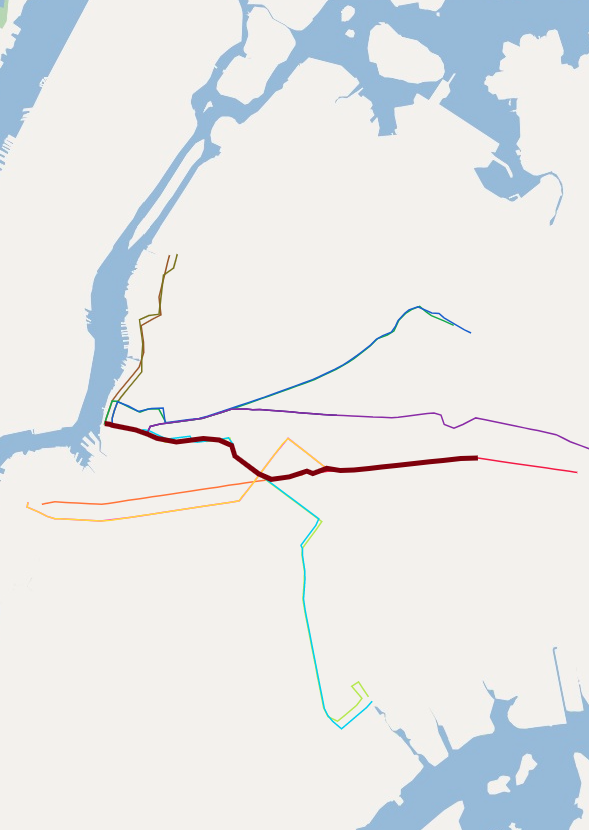}}\hspace{2.5em}
		\subfigure[\small Staten Island]{\includegraphics[height=4.4cm]{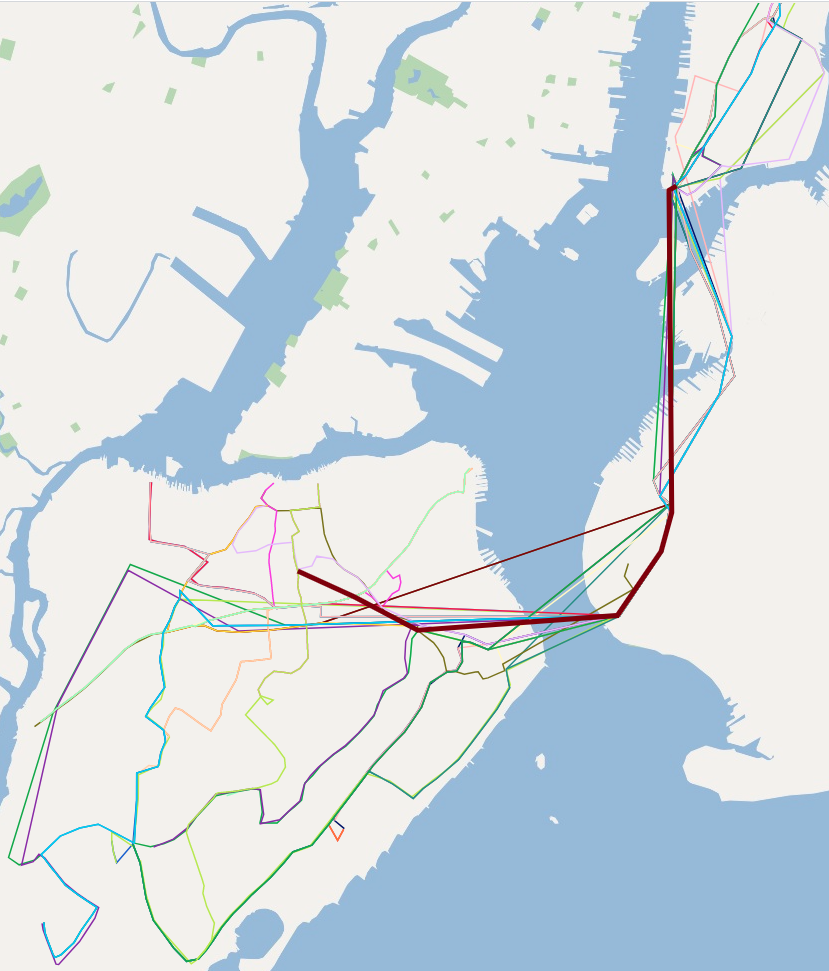}}\hspace{2.5em}
		\subfigure[\small Bronx]{\includegraphics[height=4.4cm]{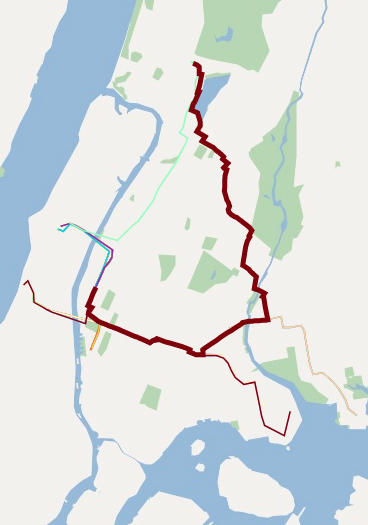}}
		\vspace{-1.5em}
		\caption{Visualization of a new bus route (bold red) and its connected existing routes ($w=0.5$).}
		\vspace{-1em}
		\label{fig:case-stuides}
	\end{minipage}
	\hspace{-0.5em}
	\begin{minipage}{0.25\textwidth}
		\centering
		\vspace{0.5em}
			\subfigure[\small $w=1$ (vk-TSP)]{\includegraphics[height=4.4cm]{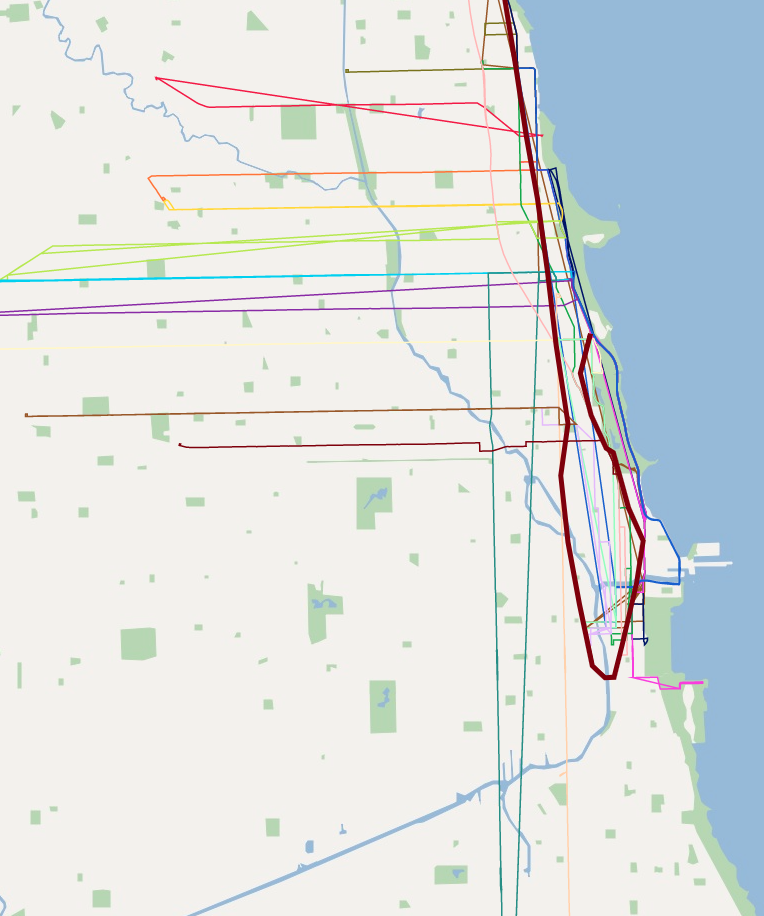}}\\\vspace{-0.7em}
		\subfigure[\small $w=0$]{\includegraphics[height=4.4cm]{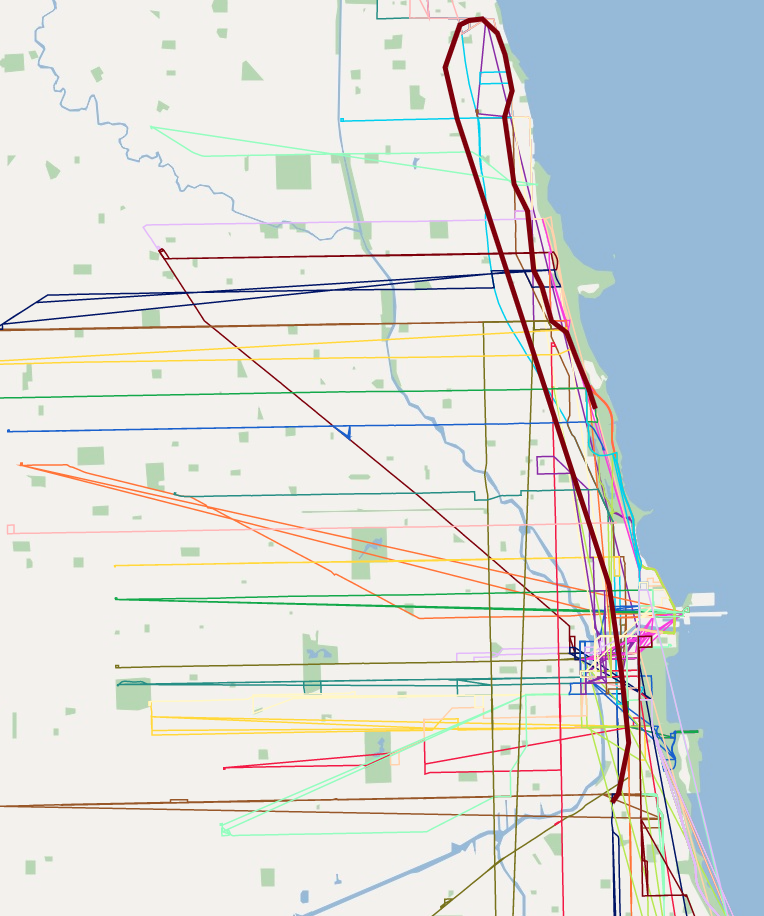}}
		\vspace{-1.5em}
		\caption{{$w=1$ \& $0$@Chicago.}}
		\vspace{-1em}
		\label{fig:weight-vis}
	\end{minipage}
\end{figure*}

For effectiveness evaluation, we first compare the increased objective value and connectivity value by our route, 
{and then introduce three metrics to measure the transfer convenience of the new transit network, as shown in Table~\ref{tab:num}.}
We also conduct a visual analysis on our taxi trajectory dataset to plan new bus routes through visualization (Figure~\ref{fig:case-stuides}). 

For efficiency evaluation, we compare the running time (Table~\ref{tab:running}) and verify the optimizations on accelerating the convergence, where we alter the parameter $k = [10, 20, \underline{30}, 40, 50]$, the weight $w=[{0.3, \underline{0.5}, 0.7}]$, the seeding number $\var{sn}=[3000,\underline{5000},7000]$, and the number of turns $\var{Tn}=[1,\underline{3},5]$. The default value is underlined.
The objective values are recorded every $100$ times, and the iteration number is set as $100,000$ in Figures~\ref{fig:precompare}-\ref{fig:efficiency1}.


For the sake of unity of objective value in Definition~\ref{def:fairbus}, we choose same values for normalization.
Specifically,
we choose top-$k$ edges and sum up their demands to set as the ${\lambda}_{max}$ for the connectivity normalization in Equation~\ref{equ:objective}. 
Similar operation applies to the demand normalization based on $L_d$.\footnote{Since $\sum_{i = 1}^k L_\lambda(i)$ and $\sum_{i = 1}^k L_d(i)$ are also used as upper bounds which are much bigger than $O_d$ and $O_\lambda$, the objective values of the final results are usually small (e.g., those in Figure~\ref{fig:precompare} \& \ref{fig:efficiency}).}

\vspace{-1em}
\begin{equation}
\label{equ:newb}
{\lambda}_{max} =  \sum_{i = 1}^k L_\lambda(i),~~~~~
{d}_{max}=  \sum_{i = 1}^k L_d(i)
\end{equation}

\begin{table*}
	\ra{1.1}
	\centering
	\caption{Effectiveness analysis of planned routes. ({normal cells: \textbf{ETA} $\vert$ \textbf{ETA-Pre} $\vert$ \textbf{vk-TSP}; gray cells: \textbf{ETA-Pre} with $w=0\ |\ 0.3\ |\ 0.7$)}}
	\vspace{-1em}
	\label{tab:num}
	\scalebox{1}{\begin{tabular}{cccccccc}
			\toprule
			\multirow{3}{*}{\textbf{City}}& \multicolumn{3}{c}{\textbf{Improvement on Defined Metrics}}&&\multicolumn{3}{c}{\textbf{Transfer Convenience Metrics}}\\ \cmidrule{2-4} \cmidrule{6-8}
			& \textbf{\makecell{\small\#New edges}} & \textbf{\makecell{\small Objective $O(\mu)$}} & \textbf{\makecell{\small Connectivity}} & & \textbf{\makecell{\small \#Transfer avoided}}&\textbf{\makecell{\small Distance ratio $\zeta(\mu)$}} &  \textbf{\makecell{\small \#Crossed routes}} \\ \midrule
			\multirow{2}{*}{Chicago}    &                               29 $\vert$ 29 $\vert$    22           &             0.22 $\vert$ 0.22 $\vert$   0.06          &                0.20 $\vert$ 0.19 $\vert$  0.05  &     & 3.02 $\vert$ 3.15  $\vert$ 2.33  & 5.35 $\vert$ 5.90 $\vert$ 5.45   &  41 $\vert$ 30 $\vert$ 25        \\		
			    &       \cellcolor{lightgray}29 $\vert$ 29 $\vert$ 29           &    \cellcolor{lightgray}    0.29 $\vert$ 0.27 $\vert$   0.16          &      \cellcolor{lightgray}          0.24 $\vert$ 0.22 $\vert$  0.15  & \cellcolor{lightgray}     &\cellcolor{lightgray} 3.43 $\vert$ 3.27  $\vert$ 2.89  & \cellcolor{lightgray}5.95 $\vert$ 5.91 $\vert$ 5.67   &\cellcolor{lightgray}60 $\vert$ 45 $\vert$ 27        \\
			Manhattan   &                        19 $\vert$ 23 $\vert$ 21          &             0.08 $\vert$ 0.07 $\vert$   0.06          &                0.17 $\vert$ 0.18 $\vert$ 0.13        & & 1.43 $\vert$ 1.40  $\vert$ 1.32& 1.86 $\vert$  1.91 $\vert$ 1.47 &     {\color{white}0}5 $\vert$ {\color{white}0}7 $\vert$ {\color{white}0}4        \\
			Queens     &                    13 $\vert$ 20  $\vert$ {\color{white}0}8          &             0.09 $\vert$ 0.09  $\vert$ 0.12           &                0.14 $\vert$ 0.17 $\vert$  0.03    &   &  4.22 $\vert$ 4.39  $\vert$ 2.76 & 1.60 $\vert$   1.59 $\vert$ 1.93&    31 $\vert$ 37  $\vert$ 22    \\
			Brooklyn    &             26 $\vert$ 26 $\vert$  {\color{white}0}6           &             0.11 $\vert$ 0.10 $\vert$  0.04            &                0.22 $\vert$ 0.23 $\vert$ 0.03    &       & 1.39 $\vert$ 1.36  $\vert$ 1.25 & 2.44 $\vert$  2.85  $\vert$ 1.16&     13 $\vert$ 17 $\vert$ {\color{white}0}5     \\
			Staten Island     &                              11 $\vert$ 11  $\vert$  {\color{white}0}6           &             0.09 $\vert$ 0.09 $\vert$  0.08           &                0.16 $\vert$ 0.16 $\vert$ 0.05   &  & 1.93 $\vert$ 1.89 $\vert$ 1.67 & 3.66 $\vert$ 3.83  $\vert$ 3.64 &         42 $\vert$ 40 $\vert$  34    \\
			Bronx     &                           21 $\vert$ 19 $\vert$   {\color{white}0}4          &             0.08 $\vert$ 0.08 $\vert$  0.01           &                0.16 $\vert$ 0.16 $\vert$  0.02    &     &  4.78 $\vert$ {\small \textbf{4.73}} $\vert$ {\small\textbf{1.60}} & 6.38 $\vert$ 7.07 $\vert$ 1.32 &      20 $\vert$ 17 $\vert$ {\color{white}0}8     \\ \bottomrule
	\end{tabular}}
\end{table*}

\subsection{Effectiveness}
\label{sec:effectiveness}
We conduct both quantitative analysis and visual analysis, showing \fairbus with \textbf{ETA} has the potential to increase connectivity and meet commuters' demand for Chicago and five boroughs of NYC.\footnote{In NYC, the bus transit system is planned independently in each borough.} 

\subsubsection{Comparisons}
We implemented two most related approaches (see the last paragraph of Section~\ref{sec:liter}) which optimize the connectivity \cite{Chan2014,Wei2014a} and demand \cite{Wangaclustering} by setting $w$ as 0 and 1, respectively. Other approaches \cite{Liu2016,Wang2018c,Chen2014a} which aim to optimize an existing bus route are essentially different from our work in term of problem formulation, hence we will not compare with them.

\myparagraph{1) Connectivity-First Approaches}
\citet{Chan2014} proposed to maximize the natural connectivity of a graph when adding $k$ new edges (not a path). 
We can use a greedy algorithm proposed by \cite{Chan2014} to generate $l<k$ new edges to maximize the natural connectivity first, connect and order them using \textit{travelling salesman search}, and then enrich the two ends by the shortest path.
However, Figure~\ref{fig:baseline-topk} shows  $10$ edges returned by this method, and they are hard to be connected as a smooth bus route. 
In addition, this greedy algorithm needs several hours to complete. 
Hence, we will not conduct further comparison with this approach.

{\myparagraph{2) Demand-First Approaches} When optimizing the trajectory-based demand alone, similar to trajectory clustering \cite{Wangaclustering}, the problem is a variant of k-TSP: \textit{maximizing the sum of edges' demands with at most $k$ new edges}. The difference with \textbf{k-TSP} is that the edges in the path should be newly connected.
	To increase the connectivity simultaneously, we set a constraint of new edges only here, as adding existing edges will not increase the network connectivity.
	We denote this baseline as \textbf{vk-TSP}.
	Note that Algorithm~\ref{alg:fairbus} applies a classical greedy method that can also work for the \textbf{vk-TSP}. To have a fair comparison when verifying the increments on connectivity in Section~\ref{sec:effectiveness}, we implemented it with the same configuration with minor changes: 1) setting $w=1$; 2) only considering new edges during the initialization and expansion.
}

\subsubsection{Results} 
\myparagraph{Analysis of Objective Values} Table~\ref{tab:num} shows the estimated connectivity and the objective value increment of new path by running \fairbus in different areas, with our two algorithms.
The connectivity values here are normalized by $\lambda_{max}$ for a better illustration.
We also compare the number of existing bus routes that can transfer to the route returned by our two methods, in order to verify whether the pre-computation sacrifies precision for efficiency.
In each cell that has three numbers, the left is returned by online computation (\textbf{ETA}), the central is returned by  pre-computation (\textbf{ETA-Pre}), {and the right is returned by our baseline (\textbf{vk-TSP})}.
It shows that \textbf{ETA-Pre} and \textbf{ETA} have similar performance,  and \textbf{ETA-Pre} is dominant in most cases.

{We compare with \textbf{vk-TSP}, which plans a route by adding new edges to maximize the demand increment only and hence should also have a considerable connectivity increment, as shown in Table~\ref{tab:num}. However, we find \textbf{ETA-Pre} has a larger connectivity increment as we append new edges with high connectivity, while \textbf{vk-TSP} appends edges with high demand but may have low connectivity increment.
We ignore the efficiency comparison as they use the same procedure in Algorithm~\ref{alg:fairbus}.}

{\myparagraph{Transfer Convenience} Since transfer convenience is one of the main performance indicators of a connected transit network \cite{Sun2016c}, we further evaluate the newly planned route's effect to the commuters along it, which are composed of an origin stop and a destination stop in the new route. For every possible trip of these commuters, we run the shortest path search on the old and new bus network, respectively. 
	We use the following three metrics widely adopted in transportation evaluation area \cite{Abdelaty2020,Zou2013}, also shown in the right part of Figure~\ref{fig:case-stuides}, and a higher value indicates more convenience.
	Then, we calculate the average value for each metric.
	
	Firstly, we calculate how many transfers are needed in the old network. Since there is no direct path like our new route, passengers need multiple transfers, e.g., it is 3.15 in Chicago.
	Secondly, for a group of trips, we calculate the ratio of shortest-path travel distance via the new bus network over that via the old bus network, i.e.,
	
	\begin{equation}
	\zeta(\mu) = \frac{1}{l(\mu)\cdot(l(\mu)-1)}\cdot\sum_{\forall O, D\in\mu}\frac{|G_r(O, D)|}{|G_r^{'}(O, D)|}
	\end{equation}
	where $O$ and $D$ are any two different stops in $\mu$, then there are $l(\mu)\cdot(l(\mu)-1)$ possible pairs, and $l(\mu)$ is the number of stops in $\mu$; $|G_r(O, D)|$ denotes the travel distance of shortest path from $O$ to $D$ in $G_r$.
	The ratio $\zeta(\mu)$ is always bigger than one as passengers can directly commute without detour anymore in the new network $G_r^{'}$, and can have shorter travel distance.
	Thirdly, we count how many existing bus routes share common stops with the newly planned one. More crossed routes mean that passengers can easily transfer and get to more destinations in the network by taking other routes.
}

{\myparagraph{Effect of Varying the Weight $w$} To investigate the effect of $w$ on the resulted routes, we set $w$ as $\{0, 0.3, 0.5, 0.7, 1\}$ on Chicago and observe how the metrics change w.r.t. $w$. 
Since \textbf{ETA-Pre} and \textbf{vk-TSP} were set with $w=0.5$ and 1, the rest results of other values (0, 0.3, 0.7) can be found from the grayed row of Table~\ref{tab:num}}.

\myparagraph{Visual Analysis} 
In Figure~\ref{fig:case-stuides},
we visualize the planned route based on our algorithms and its existing connected routes in different colors to make sure it can present a complete profile of each route.
Our newly planned routes are highlighted in bold red.
{By default, we set $w=0.5$ in Figure~\ref{fig:case-stuides}. 
To further investigate the visualized effect of $w$, we set it as $0$ and $1$ respectively and compare the results in Figure~\ref{fig:weight-vis} with those in  Figure~\ref{fig:case-stuides}(a)}.

\vspace{1em}
\noindent\textbf{\underline{\textit{Insight 1}}:}
	1) \fairbus based on our method can generate a valid path with a high objective value comparable to the one with online computation, and keep a balance between demand and connectivity. 
	{2) Compared with \textbf{vk-TSP}, our method gets a much higher connectivity increment; we further verify that new routes with higher connectivity increments can dramatically avoid transfers for commuters with direct routes, and also provide more transfer choices especially in Chicago and Bronx.}
	3) The planned routes are very smooth in the map, and it also indicates the emerging trends and valuable suggestions for new bus routing planing in each city.
	{4) With more weight on the connectivity (i.e. smaller $w$), the connectivity increases linearly, such that it becomes easier to transfer (linearly increased metrics) and meanwhile more existing routes are connected.}
	
\noindent\textbf{\underline{\textit{Insight 2}}:}
	For Chicago, \fairbus suggests that one more route should be built to connect the northwest part (Avondale) to the city, as we can see, most existing bus lines are near the lakeside.
	{Different with $w=0.5$ in Figure~\ref{fig:case-stuides}(a), we observe that the route planned with $w=1$ (considering demand only)  crosses the city and coast area to meet high demands from passengers (see Figure~\ref{fig:weight-vis}(a)), but it intersects with much fewer routes (only 25) than the planned route of $w=0$ (considering connectivity only), which connects 60 routes but crosses interior area mostly (see Figure~\ref{fig:weight-vis}(b)). The above analysis tells that a choice of $w=0.5$ can make a good trade-off.}
	
\noindent\textbf{\underline{\textit{Insight 3}}:}
	For {NYC}, more routes need to be built between Queens and Brooklyn, which will further connect more routes to Staten island.
	For Manhattan, existing subway and bus systems are very mature and connectivity increase will not be obvious, and newly planned bus routes are not necessary. This is also consistent with the fact that NYC is redesigning bus routes in the other four boroughs except for Manhattan.
	However, more routes should be planned to connect Manhattan with Staten Island which highly depends on buses, while there is only one internal subway line on the island.
	The Bronx also needs to connect north and south to form a circle from Yankee Stadium, Hunts Point Av, to Kingsbridge.


\begin{table}
	\ra{1}
	\centering
	\caption{Running time (s) comparison with increasing $k$.}
	\vspace{-1em}
	\label{tab:running}
	\scalebox{1}{\begin{tabular}{ccccc}
		\toprule
		& \textbf{{Chi-ETA}} & \textbf{\makecell{Chi-ETA\\Pre}} & \textbf{{NYC-ETA}} & \textbf{\makecell{NYC-ETA\\Pre}} \\ \midrule
		$k=10$     &         22234.21         &           55.45           &     15011.55    &        37.55         \\
		$k=20$     &         28291.92         &           76.88           &     16468.02     &        43.14         \\
		$k=30$     &         30828.44         &           82.45             &       16567.51             &     41.17                   \\
		$k=40$     &         31967.53         &            88.32            &     16671.96               &    41.13                    \\
		$k=50$     &         32435.84          &             94.14          &        16686.87            &        44.97                \\ \bottomrule
	\end{tabular}}
	\vspace{-1em}
\end{table}

\begin{figure}
	\centering
	\hspace{-1.5em}
	\subfigure[Chicago]{\includegraphics[height=2.5cm]{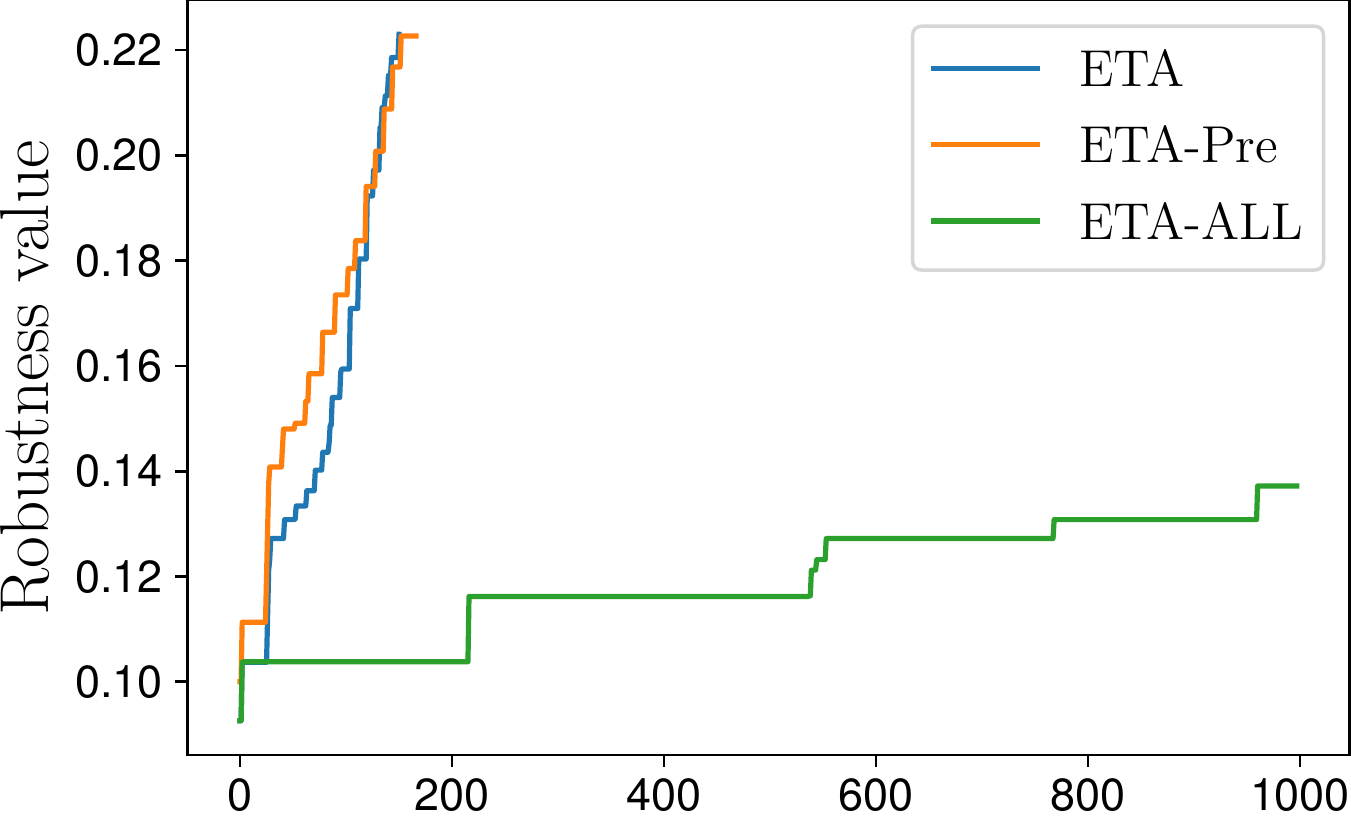}}
	\subfigure[NYC]{\includegraphics[height=2.5cm]{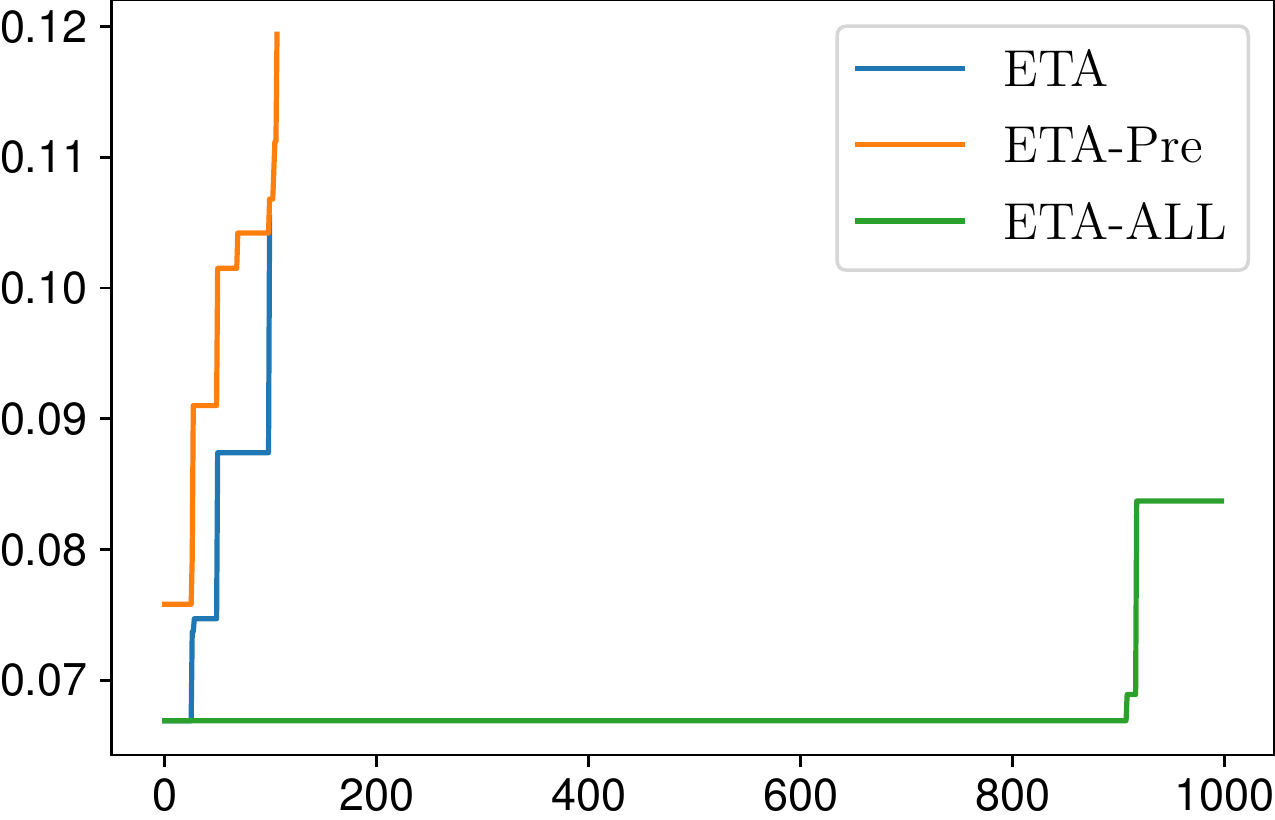}}
	\vspace{-1.5em}
	\caption{Convergence comparison of \textbf{ETA} and \textbf{ETA-Pre}.}
	\vspace{-1.5em}
	\label{fig:precompare}
\end{figure}

\begin{figure}
	\centering
	\hspace{-1.5em}
	\includegraphics[height=2.6cm]{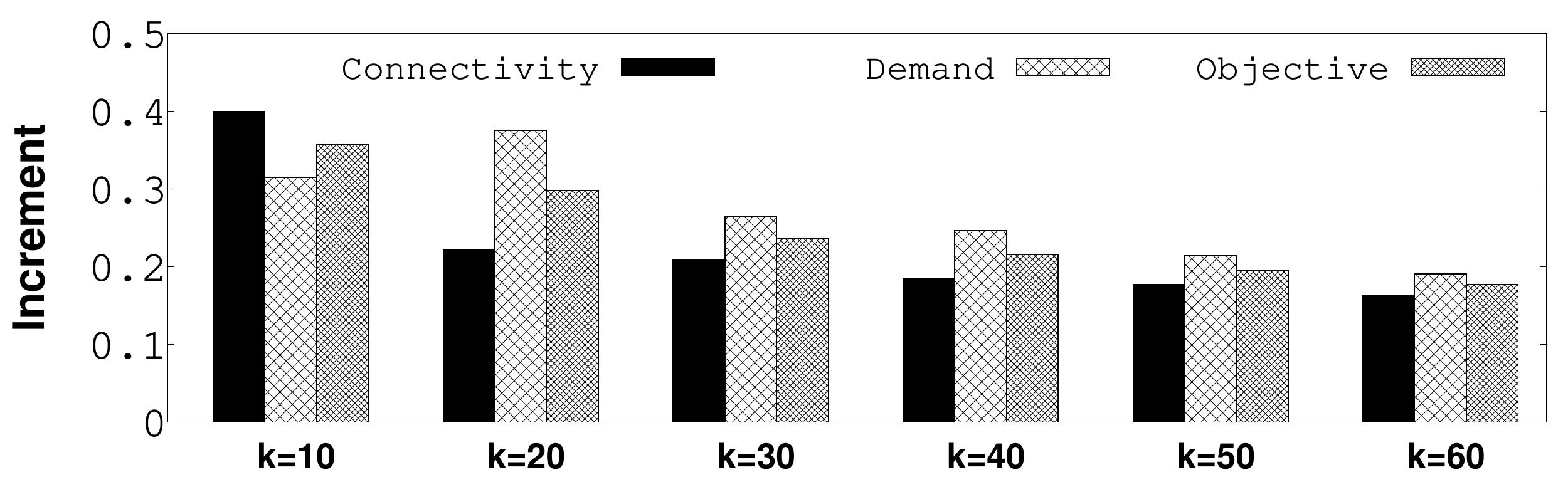}
	\vspace{-1em}
	\caption{Objective value, connectivity, and demand increments with increasing $k$.}
	\vspace{-1.5em}
	\label{fig:increment}
\end{figure}

\begin{figure*}[h]
	\centering
	\includegraphics[width=15cm]{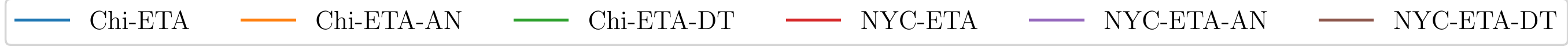}\vspace{0em}\\
	\hspace{-1em}
	\subfigure[]{\includegraphics[height=0.2\linewidth]{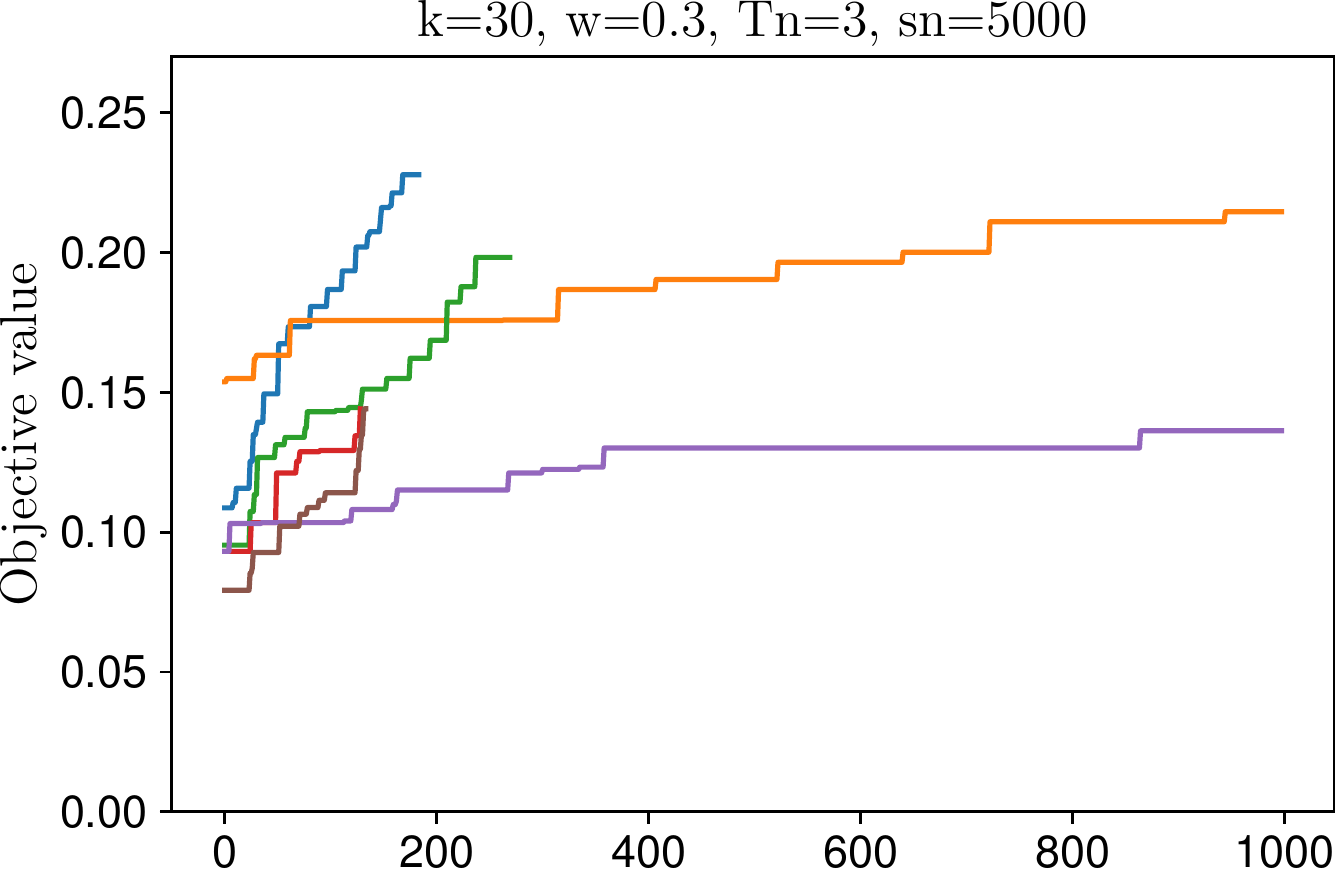}}\hspace{0.5em}
	\subfigure[]{\includegraphics[height=0.2\linewidth]{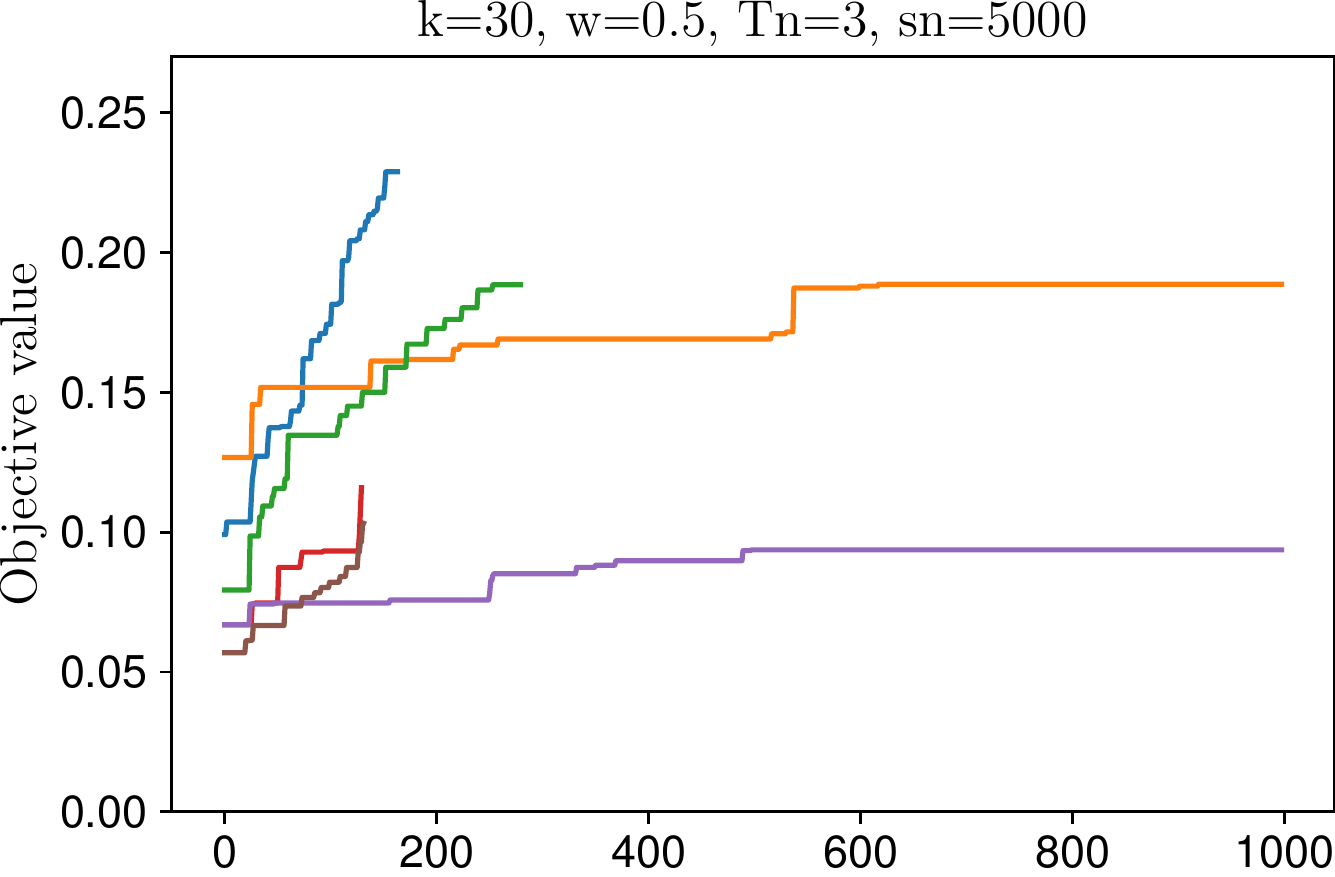}}\hspace{0.5em}
	\subfigure[]{\includegraphics[height=0.2\linewidth]{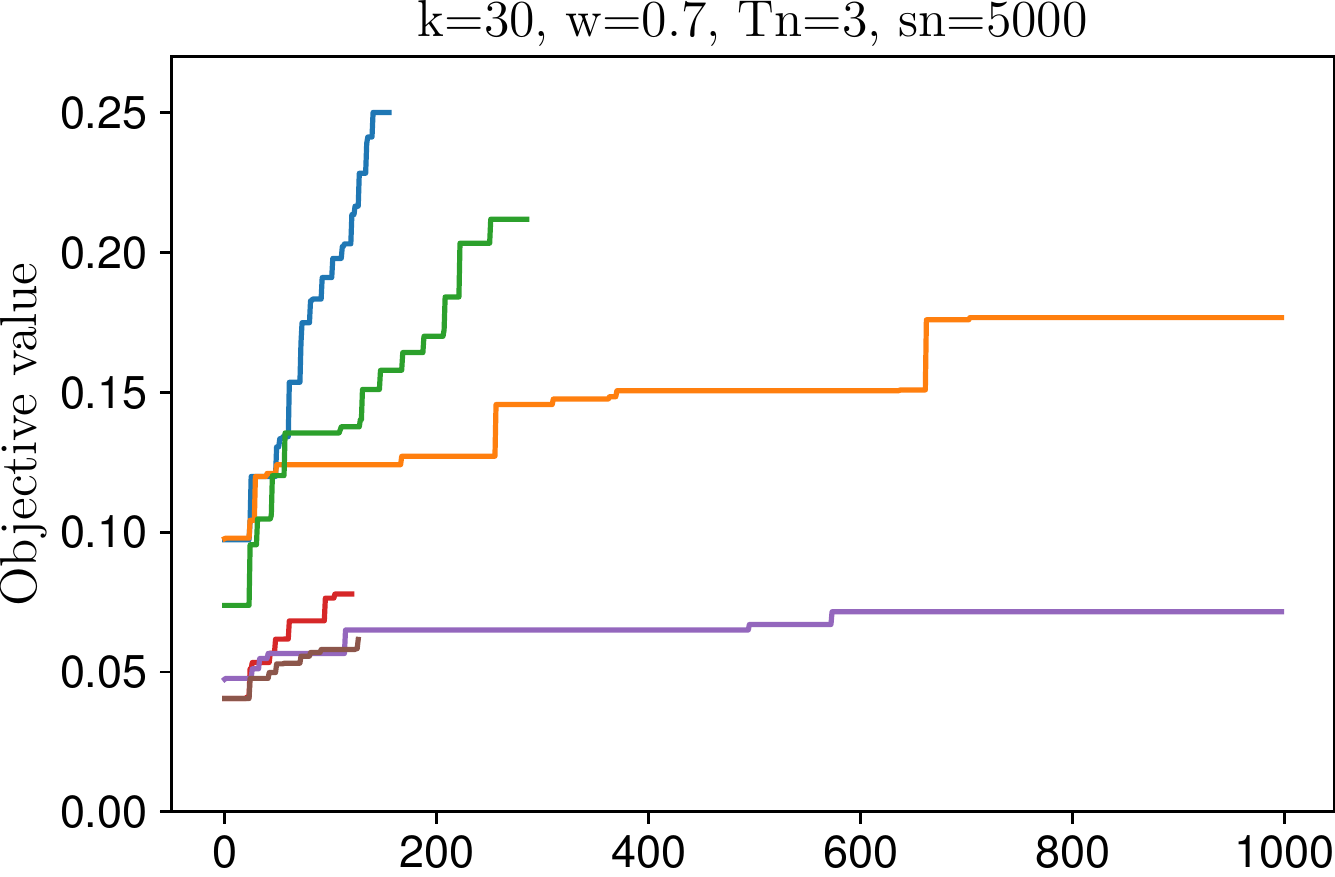}}
	\vspace{-1.5em}
	\caption{Parameter-sensitivity experiments on $w$.}
	\vspace{-1em}
	\label{fig:efficiency}
\end{figure*}

\begin{figure*}[h]
	\centering
	\hspace{-1em}
	\subfigure[]{\includegraphics[height=0.2\linewidth]{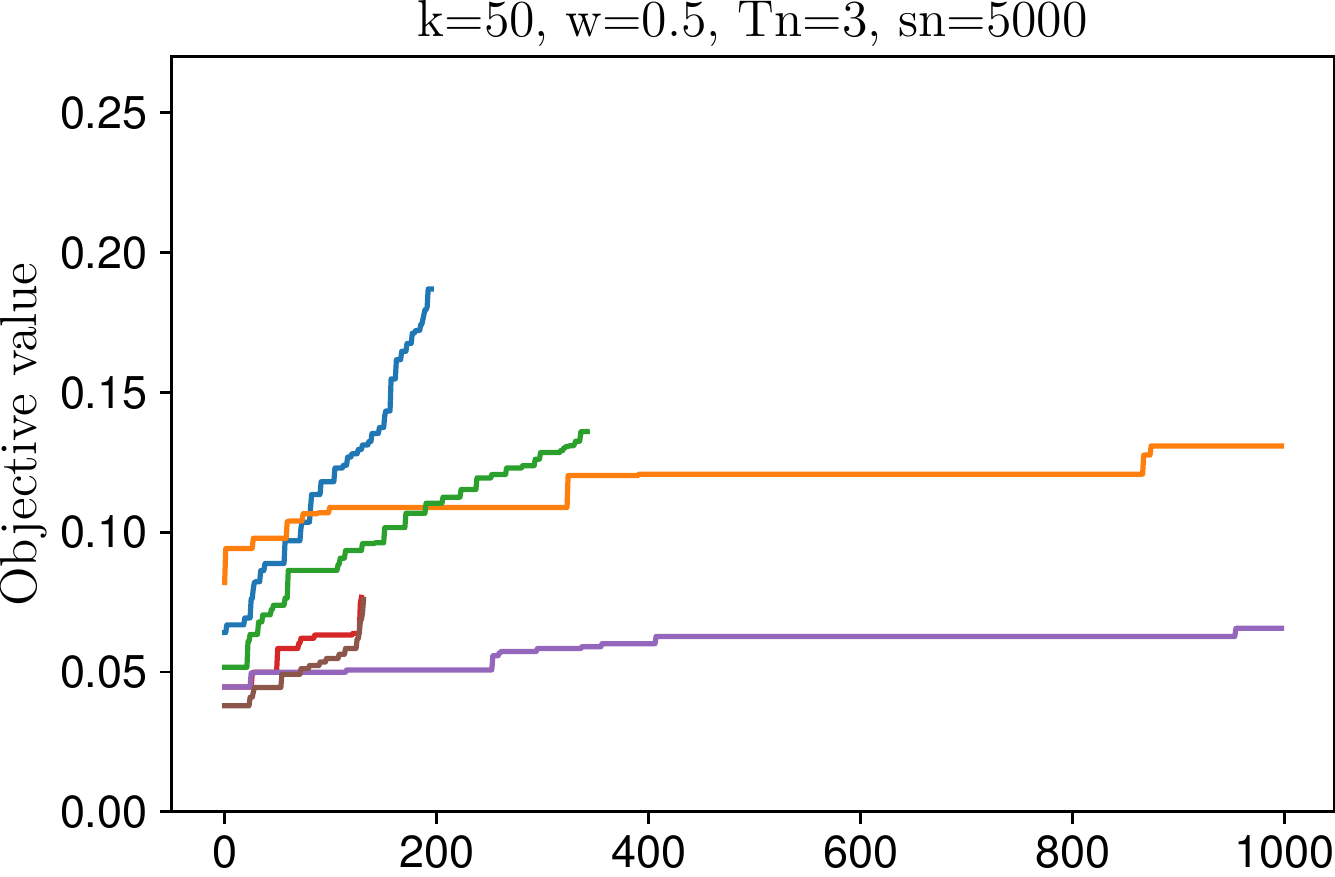}}\hspace{0.5em}
	\subfigure[]{\includegraphics[height=0.2\linewidth]{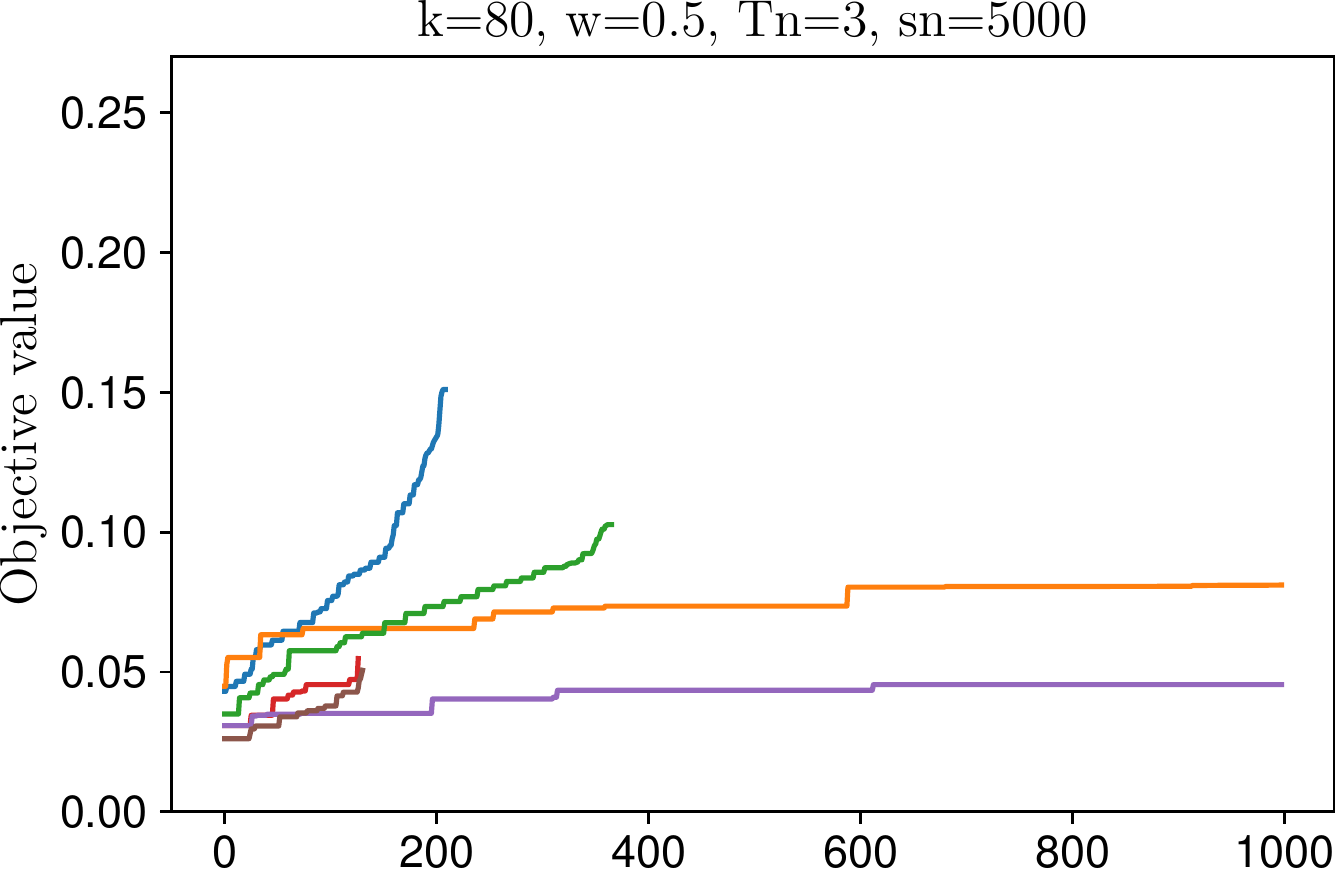}}\hspace{0.5em}
	\subfigure[]{\includegraphics[height=0.2\linewidth]{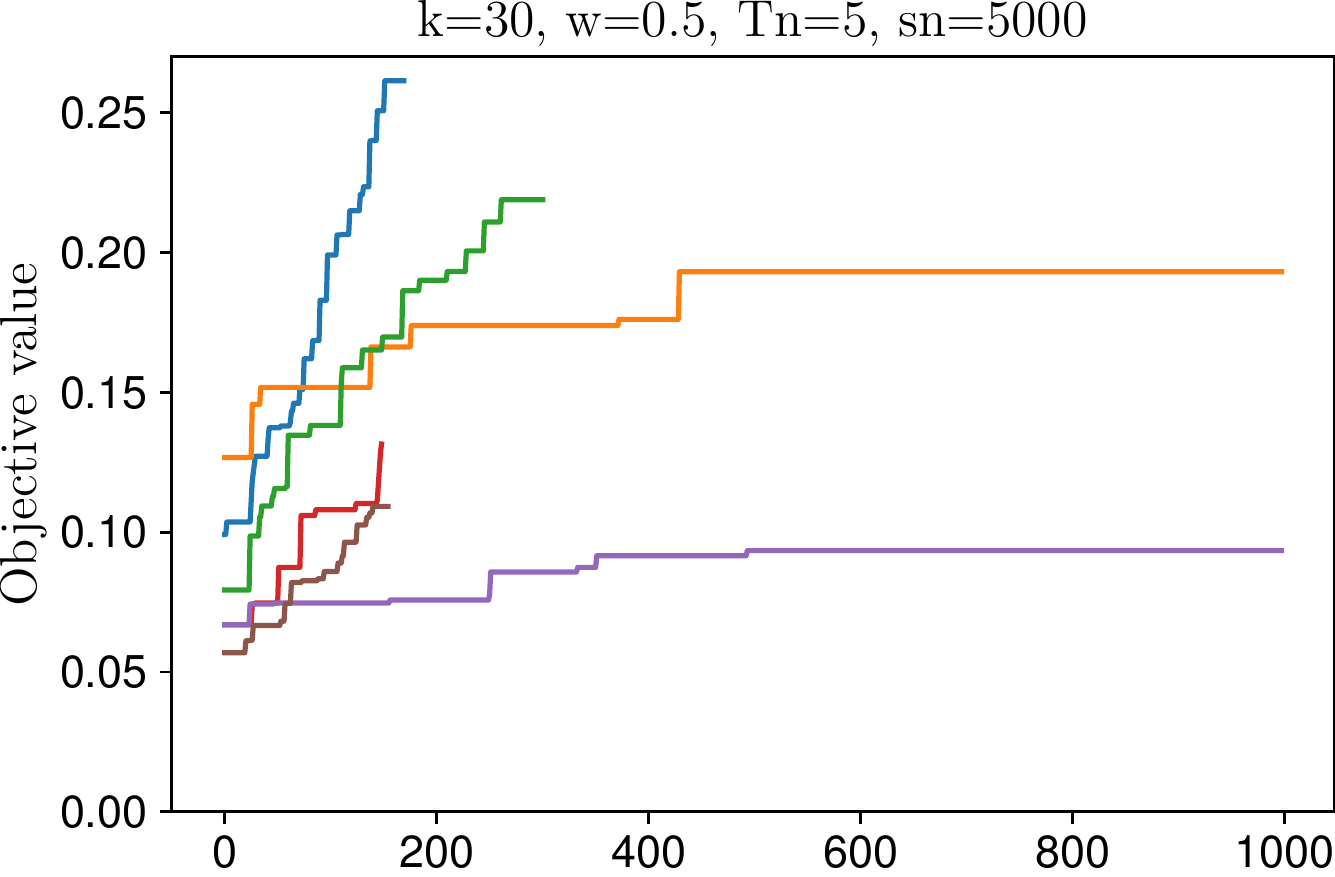}}\\ 
	\hspace{-1em}
	\subfigure[]{\includegraphics[height=0.2\linewidth]{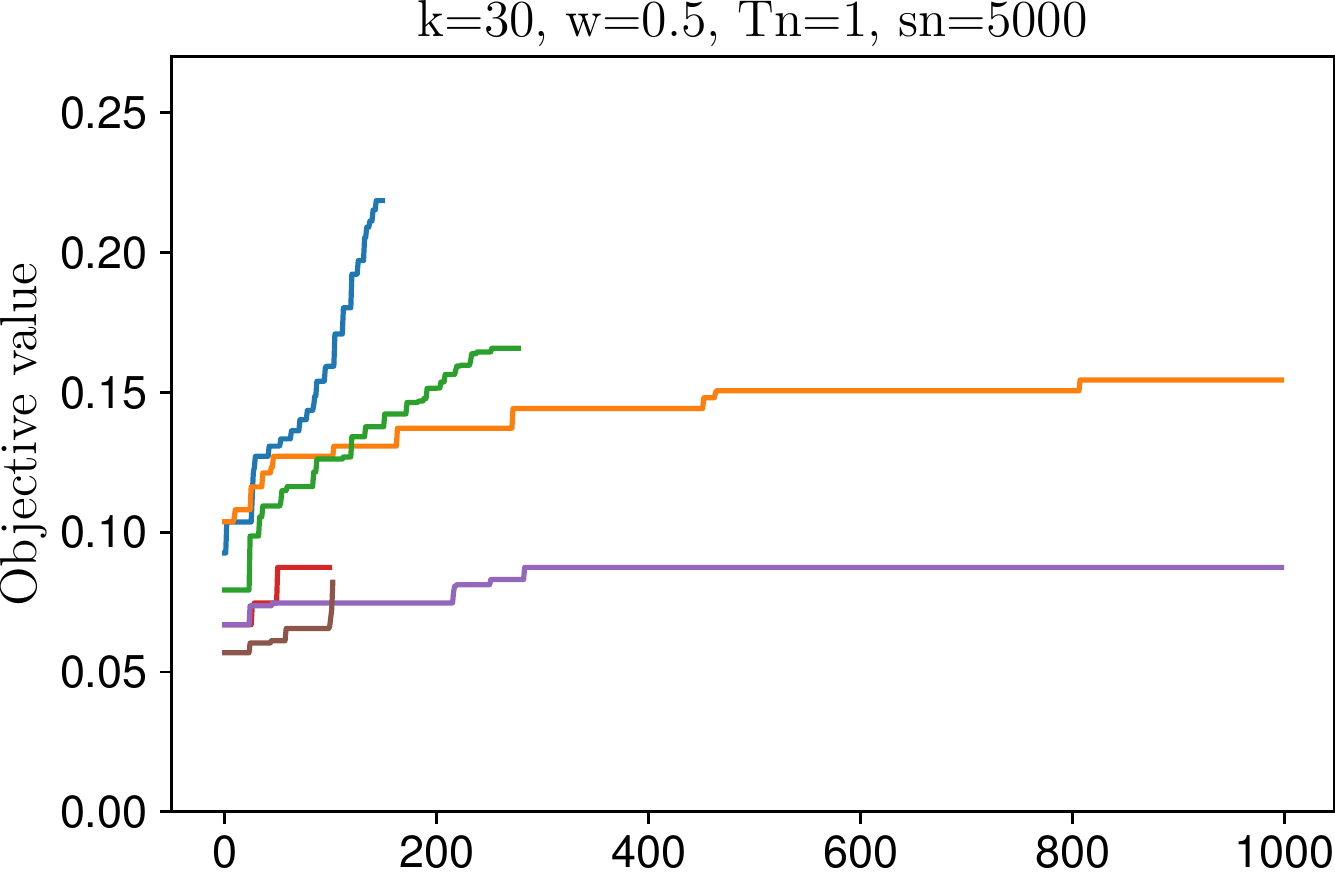}}\hspace{0.5em}
	\subfigure[]{\includegraphics[height=0.2\linewidth]{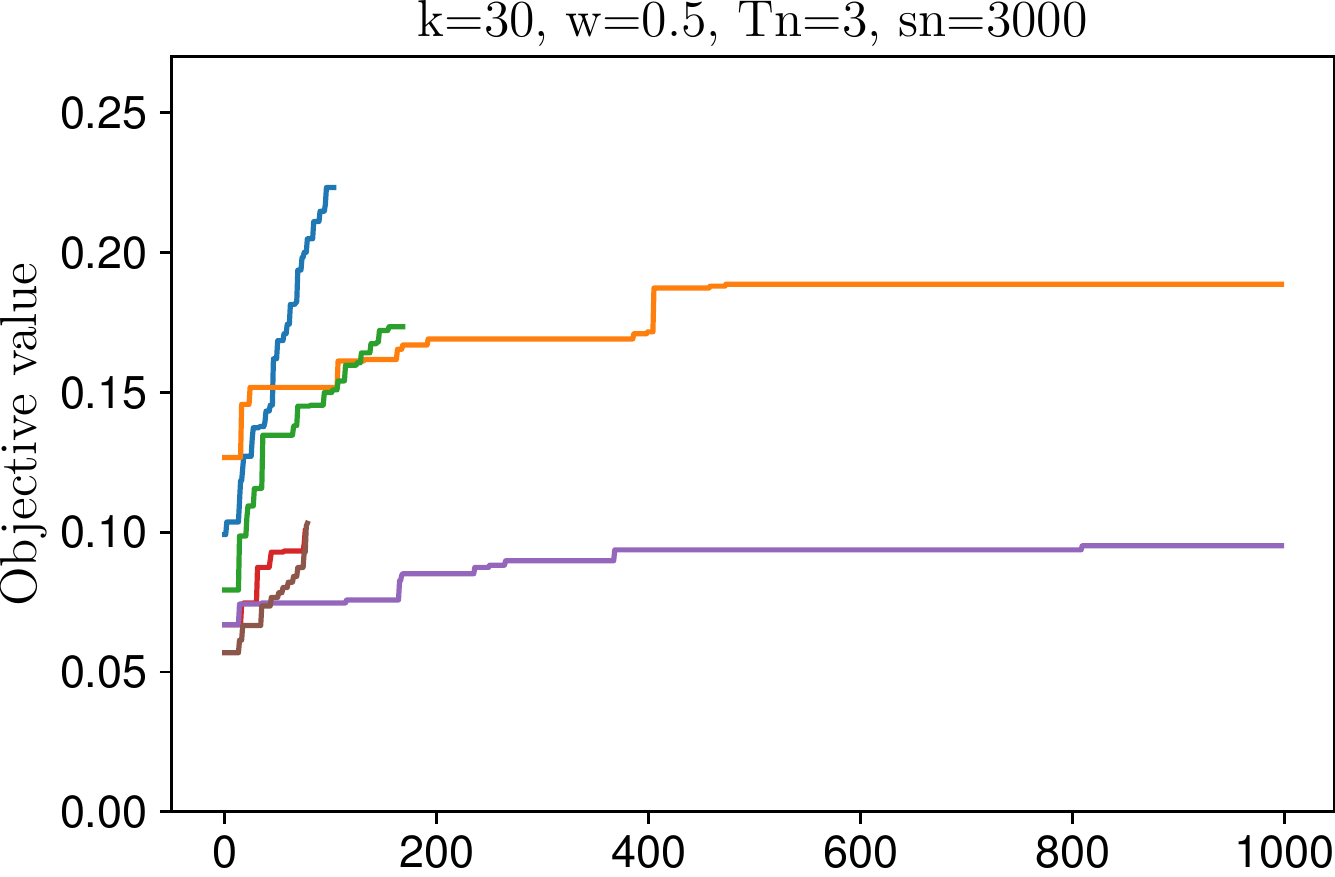}}\hspace{0.5em}
	\subfigure[]{\includegraphics[height=0.2\linewidth]{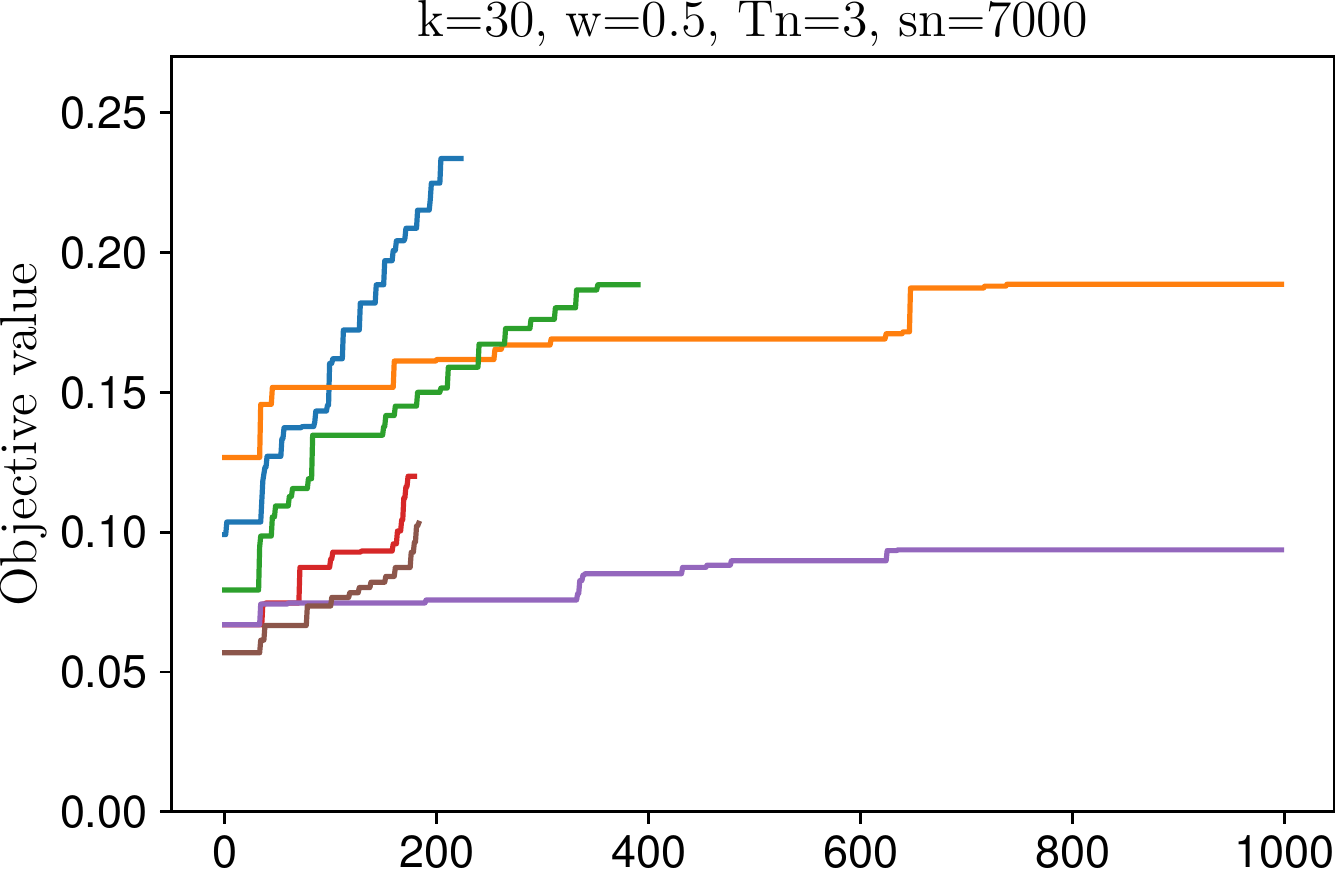}}\\
	\vspace{-1.5em}
	\caption{More parameter-sensitivity experiments on $k$, $\var{Tn}$ (number of turns), and $\var{sn}$ (seeding number).}
	\vspace{-1em}
	\label{fig:efficiency1}
\end{figure*}

\subsection{Efficiency}

\subsubsection{Comparisons}
Since \fairbus is proposed for the first time, we will mainly compare our proposed \textbf{ETA} with Lanczos and pre-computation optimizations introduced in Section~\ref{sec:computeconn} and \ref{sec:expopti}, respectively. We denote them as \textbf{ETA} and \textbf{ETA-Pre}. 
{Since ETA is based on the classical expansion-based traversal framework \cite{Wangaclustering} which takes all edges as candidate routes to start expansion, we denote it as \textbf{ETA-ALL} for comparison.}

\subsubsection{Results}
\myparagraph{Running Time of ETA}
To compare two algorithms fairly, we apply the same initialization with selective edges in Section~\ref{sec:imp} to avoid unnecessary scanning.
Table~\ref{tab:running} shows the running time using pre-computed connectivity and the \textbf{ETA} with our fast Lanczos method, by various $k$. 
We can observe that \textbf{ETA} with pre-computation (\textbf{ETA-Pre}) proposed in Section~\ref{sec:expopti} is almost 400 times faster than \textbf{ETA} with online connectivity computation, as it has computed core information for fast connectivity and bound estimation off-line, and online computation is very limited.

\myparagraph{Convergence of ETA}
Figure~\ref{fig:precompare} shows the convergence of our two methods with the increase of iteration number, where we estimate the final results of \textbf{ETA-Pre} using the Lanczos method and plot it as the last point.
We observe that \textbf{ETA-Pre} has comparable and even higher objective values due to our tight upper bounds, and initializing all edges leads to slow convergence.

\myparagraph{Parameter Sensitivity Test}
{To further verify the parameter-sensitivity of domination table (DT) and enqueueing best neighbors rather than all neighbors (AN), 
we add two more comparisons which mute them respectively. We test the sensitivity to four parameters: $k$, $w$, $\var{Tn}$, and $\var{sn}$. Also, we use \textbf{ETA-Pre} only since \textbf{ETA} is too slow to run such multiple rounds' tests.} 
%
{Figure~\ref{fig:increment} presents how the connectivity, demand and objective values change with an increasing $k$.}
{Figure~\ref{fig:efficiency} shows that despite the choice of $w$, our algorithm converges well and terminates at an early stage, as the feasibility checking has pruned all the candidate paths and the queue is empty. 
Figure~\ref{fig:efficiency1} shows the sensitivity result on the rest parameters. 

We find that the objective values drop with an increase of $k$ (see Figure~\ref{fig:efficiency}, Figure~\ref{fig:efficiency}(b), and Figure~\ref{fig:efficiency1}(a)(b)), because our normalization values $d_{max}$ and $\lambda_{max}$, which are related to $k$ in Equation~\ref{equ:newb}, also rise but with a bigger increase rate than $O_d$ and $O_\lambda$.
{Figure~\ref{fig:efficiency} also indicates that our weight parameter keeps a good balance between connectivity and demand, and the demand slightly dominates the connectivity when $k$ is large. This is because more existing edges are inserted into the route when no more feasible new edges can be found.}
For other parameters like $w$ (see Figure~\ref{fig:efficiency}), $\var{Tn}$ (see Figure~\ref{fig:efficiency}(b) and Figure~\ref{fig:efficiency1}(c)(d)), and $\var{sn}$ (see Figure~\ref{fig:efficiency}(b) and Figure~\ref{fig:efficiency1}(e)(f)), none of them has much impact to the convergence and efficiency.
} 

\vspace{0.5em}
\noindent\textbf{\underline{\textit{Insight 4}}:}
	1) Our \textbf{ETA-Pre} (i.e. with pre-computation) can converge quickly and is robust to various parameter settings. It also returns a highly similar objective score to the one with online connectivity computation while the latter is much slower. 
	2) Both the best-neighbor-only optimization strategy and the domination table optimization strategy can effectively prune candidates. 
	3) Pre-computation can be done in hours but it contributes to high performance for interactive route planning \cite{Weng2020a}.

\section{Conclusions}

We investigated a public transport route planning problem \fairbus, which aims to plan a bus route to improve the connectivity of the transit network and also to meet the demand of commuters. We formulated \fairbus as an optimization problem and proposed a practical heuristic method to solve it. To avoid computationally expansive matrix operations, we used the Lanczos method to estimate the natural connectivity of transit network with bounded error. We derived upper bounds on the objective values when adding edges, and used the derived upper bounds to select edges for greedy expansion. Our experiments showed that \fairbus could plan effective routes in two of the US's most complicated bus transit systems.

In future, we will investigate how to update the connectivity efficiently in the pre-computation stage based on perturbation theory, and use our derived upper bounds to solve existing and new network connectivity optimization problems \cite{Chan2014,Chen2018d}.
For small-scale cities that do not have sophisticated transit systems, 
the optimal site selection for deploying new bus stops based on trajectories and connectivity will be another interesting direction for future research.

\begin{acks}
Zhifeng Bao is supported in part by ARC DP200102611, DP180102050, and a Google Faculty Award.
\end{acks}
\balance




\newpage

\bibliographystyle{ACM-Reference-Format}
\bibliography{musco.bib,library.bib,url.bib}  

\appendixpage

\appendix
\section{Proof of Lemmas}

\subsection{Proof of Lemma~\ref{lem:general_edge_additions}}

\begin{proof}
	Let $\bm{A}$ be the adjacency matrix of our original transit network $G_r$ and let $\bm{A}'$ be the adjacency matrix of the updated network $G_r'$, which is obtained by adding $k$ edges.
	Let $\bm{K} = \bm{A}'-\bm{A}$. Let $\lambda_1 \geq \ldots \geq \lambda_n$, $\lambda_1' \geq \ldots \geq \lambda_n'$, and $\sigma_1 \geq \ldots \geq \sigma_n$ be the eigenvalues of $\bm{A}$, $\bm{A}'$, and $\bm{K}$, respectively.
	
	Let $\tr(\cdot)$ denote the matrix trace, we have $\lambda(G_r') = \ln\left(\frac{1}{n}\tr(e^{\bm{A}'})\right)$. Then it suffices to upper bound $\tr(e^{\bm{A}'})$. To do so, we apply the Golden–Thompson inequality: $\tr(e^{\bm{A}'}) = \tr(e^{\bm{A}+\bm{K}}) \leq \tr(e^{\bm{A}}e^{\bm{K}})$. Next, we apply a trace inequality of Lassere \cite{lasserre1995trace} to bound
	$\tr(e^{\bm{A}}e^{\bm{K}}) \leq \sum_{i=1}^ne^{\lambda_i}e^{\sigma_i}$. Since $\bm{K}$ is a graph adjacency matrix with at most $2k$ nodes, it has a rank of at most $2k$, so we have $\sum_{i=1}^n e^{\lambda_i} e^{\sigma_i} \leq \sum_{i=2k+1}^n e^{\lambda_i} + e^{\lambda_1}\sum_{i=1}^{2k} e^{\sigma_i}$. Since this expression is maximized exactly when $\bm{K}$ is chosen to maximize the Estrada index $\sum_{i=1}^{2k} e^{\sigma_i}$, we can apply the upper bound of \citet{DeLaPena2007} to obtain $\sum_{i=1}^{2k} e^{\sigma_i} \leq 2k - 1 +e^{\sqrt{2k}}$.
	
	Our final result is that $\tr(e^{\bm{A}'}) \leq \tr(e^{\bm{A}}) - \sum_{i=1}^{2k} e^{\lambda_i} + e^{\lambda_1}[2k - 1 +e^{\sqrt{2k}}]$, which gives the bound of the lemma after renormalizing and taking a log.
\end{proof}

\subsection{Proof of Lemma~\ref{lem:path_edge_additions}}

\begin{proof}
	For $i = 1,\ldots, n$, let $\Delta_i = \lambda_i' - \lambda_i$. We have that:
	\begin{align}
		\label{eq:to_max}
		e^{\lambda(G_r')} = e^{\lambda(G_r)} + \frac{1}{n}\sum_{i=1}^{n} (e^{\Delta_i} - 1)e^{\lambda_i}. 
	\end{align}
	We are going to choose $\Delta_1^*, \ldots, \Delta_n^*$ to maximize this expression given the constraints of implied by Equation \ref{eq:ky_fan}:
	\begin{align*}
		&\text{For all $q = 1,\ldots, n$,} & \sum_{i=1}^q \Delta_i &\leq \sum_{i=1}^q \sigma_i.
	\end{align*}
	It is clear that any solution which maximizes Equation~\ref{eq:to_max} under these constraints must set $\Delta_q^* = \sum_{i=1}^q \sigma_i - \sum_{i=1}^{q-1} \sigma_i = \sigma_q$. 
	\vspace{1em}
	
	The non-zero eigenvalues of $\bm{K} = \bm{A}'-\bm{A}$ are simply the well-known eigenvalues of an unweighted simple path graph, which are equal to $2\cos\left(\frac{i\pi}{k+2}\right)$ for $i = 1,\ldots,k+1$. The remaining $n-k-1$ eigenvalues of $\bm{K}$ are equal to $0$.
	Noting that only the first $m = \lfloor\frac{k+1}{2}\rfloor$ path graph eigenvalues are positive, we immediately have that $\Delta_1^*, \ldots \Delta_{m}^* = 2\cos\left(\frac{1\pi}{k+2}\right), \ldots, 2\cos\left(\frac{m\pi}{k+2}\right)$, $ \Delta_{m+1}^*, \ldots,\Delta_{n-k+m}^* = 0$, and $\Delta_{n-k+m+1}^*,\ldots, \Delta_{n}^* =\allowbreak 2\cos\left(\frac{(m+1)\pi}{k+2}\right), \ldots, 2\cos\left(\frac{(k+1)\pi}{k+2}\right)$.
	
	The lemma follows by noting that $(e^\Delta_i - 1) \leq 1$ for all $\Delta_i \leq 0$.
\end{proof}

\end{document}